# Physics-aware neural networks enable robust and full atomic structure determination via low-dose atomic electron tomography

Yao Zhang[1], Lanyi Cao[1], Zhen Sun[1], Jihan Zhou[1]*

[1]Beijing National Laboratory for Molecular Sciences, Center for Integrated Spectroscopy, College of Chemistry and Molecular Engineering, Peking University, Beijing 100871, China

*Correspondence and requests for materials should be addressed to J. Z. (email: jhzhou@pku.edu.cn)

**Abstract**

Atomic electron tomography (AET) determines the three-dimensional (3D) coordinates and chemical identities of individual atoms from a series of scanning transmission electron microscopy images taken at different tilt angles. However, under the low-dose conditions required to mitigate beam damage, the reduced signal-to-noise ratio forces a trade-off among accuracy, robustness, and throughput, which ultimately limits the broader application of AET. Here, we introduce a physics-aware, two-stage neural networks (PANN) that incorporates physical constraints throughout its workflow to achieve accurate AET under low-dose imaging. First, a global-local 3D ResUNet refines the initial reconstruction and corrects geometric distortions in the volume. Second, the local density around each identified atom is encoded using 3D Zernike moments. These feature descriptors, along with the atomic coordinates are then processed by a graph-attention Transformer to classify the elemental species. We benchmark the PANN workflow using a dataset of 42,588 reconstructed volumes, covering diverse noise models, materials morphologies, and dose settings. Under low-dose conditions, PANN significantly improves performance, reducing the atomic coordinates error and leading to an increase in the atomic recovery rate. The framework's performance on experimental lose-dose AET data across nanoparticles of varying morphology and composition demonstrate robust generalization. We anticipate this approach will extend the applicability of AET, particularly in investigating materials sensitive to electron dose or chemical state, including halide perovskites, zeolite, and quantum dot.

**Main**

Mapping the three-dimensional (3D) atomic positions in nanostructures is fundamental to understanding and controlling their properties. This enables a first-principles approach to pivotal phenomena including catalysis[1–6], ferroelectricity[7–10], atomic disorder in glasses[11–15], and defect physics[16–20]. Atomic electron tomography (AET) determines the 3D atomic arrangement in non-periodic structures by combining aberration-corrected scanning transmission electron microscopy (STEM) with computational reconstruction[21–25]. This technique provides direct access to critical structural features such as dislocations[20], interfacial mixing[26] and local coordination environments in heterogeneous materials[1]. However, the atomic positional and elemental precision of AET is fundamentally limited by dose-constrained imaging conditions and various noise factors, such as misalignment and missing wedge[27,28]. These errors propagate into and degrade the reliability of subsequent quantitative structural analysis.

Several methods aim to enhance the precision of atomic coordinates and elemental classification in AET. For coordinate refinement, a UNet-based approach was developed to optimize the reconstruction volumes of Pt nanoparticles[29]. However, it struggles with generalization beyond single-element, single-crystalline systems and could introduce artifacts (e.g., ghost atoms). For element classification, an unbiased iterative algorithm evaluates atomic species by testing the elemental species and evaluating the consistency of the reconstruction[23,26]. However, this process is computationally expensive and sensitive to parameter choices. Critically, both methods lack rigorous validation under practical low-dose conditions, failing to establish a reliable benchmark. Therefore, a robust, efficient, and generalizable workflow for low-dose AET remains essential.

Recent advances in machine learning have substantially improved the fidelity, robustness, and automation of 3D volume processing in electron tomography, particularly for denoising, missing-wedge correction, and volume reconstruction[30–33]. In cryo-electron microscopy (cryo-EM), software packages such as RELION[34] and CryoSPARC[35] have automated workflows and reduced manual intervention. Architectures such as ResUNet[36–42], global-local designs[43,44], and graph attention networks (GAT)[45,46] further enhance refinement and classification performance. The integration of physical constraints into combined neural networks could further advance automated electron tomography (AET), extending its efficiency and applicability to complex material systems[31,32].

Here, we present Physics-Aware Neural Network (PANN), a two-stage, post-reconstruction refinement workflow for low-dose AET. PANN leverages a Global-Local AET ResUNet-3D (GLARE) model to correct volume distortions by global-distortion-aware volume refinement, followed by a Distance-Aware Set Transformer (DAST) model that incorporates physical constraints for precise element classification. GLARE integrates feature-wise linear modulation (FiLM) layers[43,47] into a multi-scale ResUNet architecture. This enables the network to learn and apply global contextual features, effectively correcting the global distortions across the entire reconstructed volume. After training the GLARE and DAST networks on 17,340 simulated data samples with varying noise, morphology, and composition, the framework achieves a marked improvement. Under the typical noise conditions, the atomic coordinate error is reduced from ≈ 0.24 Å to ≈ 0.10 Å, and the overall atomic recovery rate (considering both atomic coordinates and atomic element species) increases from ≈ 93% to ≈ 99%. Moreover, the PANN workflow allows AET data acquisition at lower doses and higher-noise levels, thereby broadening the applicability of AET to a wider range of materials. We assessed this capability using 42,588 simulated volumes with varying noise. A practical demonstration on an experimental low-dose tilt series ($9 \times 10^4$ e$^-$ Å$^{-2}$) showed that PANN refinement recovers reconstruction quality to a level matching that of normal-dose data ($6 \times 10^5$ e$^-$ Å$^{-2}$). Through additional fine-tuning, on multi-slice simulated datasets, our method enabled atomic-resolution reconstruction of beam-sensitive systems, including perovskites, at electron doses as low as 5,000 e$^-$ Å$^{-2}$. We anticipate that PANN workflow will transform AET into a robust and quantitative technique for low-dose 3D atomic mapping. This paves the way for reliable, efficient, and interpretable atomic-scale analysis across diverse material science applications.

## Results

### Overall framework

The two-stage PANN framework is schematically illustrated in Figure 1. First, we adopt the Global-Local AET ResUNet 3D (GLARE) model that refines and geometrically corrects the input low-dose reconstruction (Fig. 1a). A key component is a global context branch, which provides summary features that modulate the network via feature-wise linear modulation (FiLM) layers for the global refinement (Supplementary Fig. 1). Then we locate atomic coordinates from the refined volume using the Roger polynomial atom tracing algorithm. Second, we use a Distance-Aware Set Transformer (DAST) architecture (Supplementary Fig. 2) to classify the chemical species of each atom. For this purpose, we extract a local feature descriptor by computing 3D Zernike moments from the density surrounding each atom. These Zernike coefficients, together with the coordinates are fed into the element determination neural network (Fig. 1b); full details are provided in Methods.

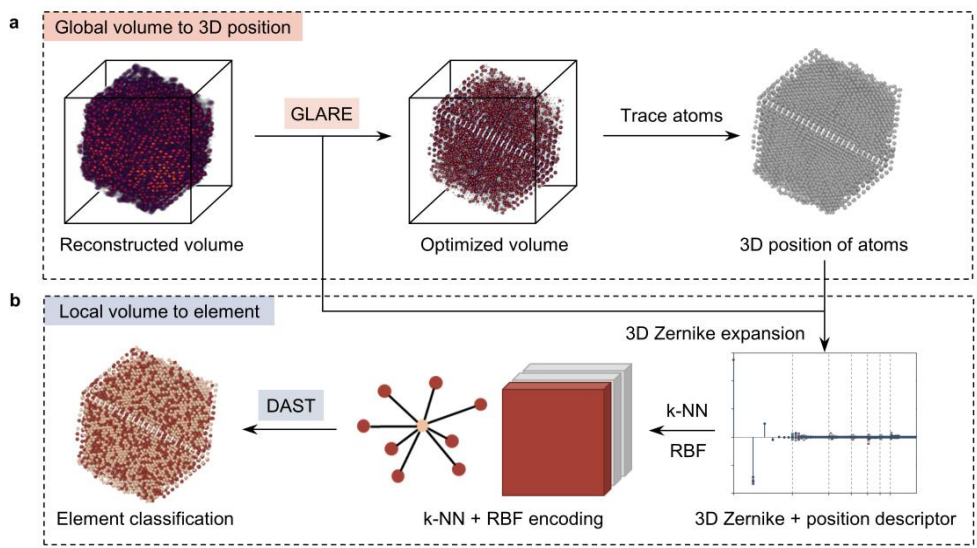

**Fig. 1 | The overall framework of PANN workflow. a,** Using GLARE model to refine the raw reconstructed volume. Subsequently, the 3D atomic coordinates are obtained from the optimized volume using Roger polynomial tracing algorithm. **b,** After the atoms traced, the 3D Zernike expansion coefficients are obtained by both atomic coordinates and the raw reconstructed volume. Then, the expansion coefficients and the atomic coordinates are encoded and embedded into a graph representation, which is subsequently processed by the DAST network for element classification.

## Performance of the GLARE network

Figure 2 shows the performance of the GLARE network for the volume refinement. The model architecture, dataset construction, and training process are provided in Methods. Both training and validation loss curves of the GLARE model decrease steadily from ≈ 0.08 to ≈ 0.035 over 145 epochs (Fig 2a), indicating stable convergence without overfitting. For comparisons, we benchmark GLARE against two reference models: a standard ResUNet[36] and a Shifted window Transformer (Swin)[48] trained under identical conditions.

We compare the reconstruction performance of GLARE against the reference methods (Fig. 2b,c). The F1-score distributions for each method are evaluated on a shared test set of 500 simulated volumes (Methods). This test set simulates realistic AET conditions by incorporating key error factors: angular/spatial misalignment, varying electron dose, the missing wedge, background intensity, Gaussian noise, aberration, defocus, sample morphology, chemical composition, and pixel size. The parameter distributions of these errors are set according to typical AET experimental conditions[20,23,26]. The test results show a clear performance hierarchy. Atoms traced directly from the raw reconstructed yield a median F1 score of 0.9666 (baseline). Refinement by ResUNet, Swin, and GLARE networks enhances the median F1 scores to 0.9848, 0.9955, and 0.9978, respectively. Moreover, the distribution range of the F1 score metrics decreases sequentially from the baseline (direct tracing) through each refinement method, highlighting the superior stability of the GLARE-refined volumes. Figure 2c shows the root mean square deviation (RMSD) distributions for the traced atomic coordinates, calculated only for atoms correctly matched to the ground truth. The median RMSD decreases from of 0.2444 Å (direct tracing from the raw volume) to 0.1907 Å

(ResUNet), 0.1670 Å (Swin), and 0.1002 Å (GLARE). Correspondingly, the dispersion of the RMSD distribution narrows progressively across these methods. These results indicate that GLARE demonstrates superior refinement capability, significantly surpassing both the baseline and other reference methods in accuracy and precision. Critically, GLARE was trained solely for volume refinement without exposure to atom tracing tasks. This design avoids the possibility that the observed improvements in RMSD and F1 score stem from overfitting to the tracing objective, confirming that the gains are attributable to genuine enhancement of the reconstruction quality.

Figure 2d,e shows the influence of key experimental error factors on reconstruction metrics under higher-noise conditions. We focus on angular misalignment, electron dose of each projection, the missing wedge, and background intensity, given their direct impact on both final reconstruction quality and the electron dose efficiency of AET data acquisition. To evaluate robustness under different noise levels, we tested both direct tracing and GLARE-refined tracing on 507 samples with diverse morphologies and compositions. The test introduced error factors at levels exceeding typical AET conditions, simulating a high-stress scenario for reconstruction algorithms. The mean F1 score and RMSD varied with different error factors respectively (Fig. 2d,e), where the shaded regions represent the 99.9% confidence intervals of the corresponding metrics. The dashed lines show the average metric values of the F1 score and RMSD from direct atomic tracing under typical-noise conditions (Fig. 2b,c), providing a benchmark for comparison.

The F1 score, defined as the harmonic mean of precision and recall, quantifies the overall performance of atomic localization (see Supplementary Fig. 3 for individual precision and recall metrics). As each error increases, the F1 score declines (Fig. 2d), indicating reduced agreement between the reconstructed and ground-truth atomic structures; meanwhile, the performance gap between GLARE refined and directly traced results widens with larger error (Supplementary Fig. 4), demonstrating that GLARE substantially slows the degradation of localization accuracy under deteriorating conditions. GLARE refinement yields consistently narrower confidence intervals for F1 scores than direct tracing across all error conditions. As errors increase, while RMSD (atoms matched to ground-truth atoms within 1 Å; Methods) increases with larger errors, indicating reduced positional accuracy (Fig. 2e). Critically, GLARE lowers RMSD by at least 0.1 Å and never underperforms direct tracing within the tested range (Supplementary Fig. 4), with both methods showing comparable confidence intervals. Under higher-noise conditions (misalignment $\leq 2.2°$, dose $\geq 10$ pA, missing wedge $\leq 65°$, and normalized background intensity $\leq 0.20–0.25$), GLARE refined tracing results achieve F1 scores surpassing the typical-noise direct-tracing baseline (Fig. 2d, dashed lines; F1 > 0.9666). Direct tracing under the same high-noise conditions, however, fails severely (Fig. 2d,e, blue lines; average F1 < 0.85 and RMSD > 0.3 Å). This demonstrates that GLARE provides robustness necessary for high-fidelity atomic analysis in challenging, dose-limited regimes. Detailed sample-by-sample comparisons (Supplementary Fig. 5) show the systematic improvement achieved through GLARE optimization over direct tracing.

We further evaluated GLARE on nanoparticle volumes reconstructed from multi-slice simulated tilt series. To mimic realistic experimental errors, controlled misalignments were introduced into the tilt-angle series. We then compared the atomic coordinates directly traced from the raw reconstruction, the UNet-refined volumes, and the GLARE-refined volumes with the ground truth in terms of RMSD and F1 score (Supplementary Fig. 6). GLARE maintained stable performance even in the presence of angular misalignment ($\leq 2°$), demonstrating that its improvement is not merely visual, but reflects a more physically faithful reconstruction of the underlying atomic structure.

Next, we demonstrate GLARE's performance on experimental AET data (Fig. 2f-m). One nanoparticle comprises ≈8,800 atoms, and its tilt-series data were collected under typical AET acquisition parameters (Supplementary Table 1). The same cross-sectional slices (1.37 Å thick) of reconstructed volumes from a Pd@Pt core-shell nanoparticle before and after GLARE refinement show the reconstruction quality (Fig. 2f & g). The raw reconstruction suffers from pronounced noise and reconstruction artifacts (Fig. 2f), which introduce ambiguity in atomic positions. GLARE refinement removes these distortions (Fig. 2g), yielding a denoised volume with enhanced clarity for reliable downstream analysis. Structural analyses confirm the improvement on this particle. GLARE-refined structure exhibits a pair distribution function (PDF) with sharper features and a narrowed bond-length distribution (Supplementary Fig. 7a-c), both signatures of enhanced structural coherence. These results collectively indicate a structurally more rational and physically plausible atomic model. Furthermore, GLARE shows powerful enhancement on more complicated polycrystalline Pd@Pt core-shell nanoparticle with ≈5400 atoms (Supplementary Table 2). The refined reconstruction volume exhibits enhanced local spherical symmetry (Methods) compared to the raw volume (Fig. 2h & k), where different colors represent the local spherical symmetry values of the corresponding atoms, indicating substantially improved reconstruction quality. Magnified views of different crystalline domains reveal clearer atomic features. In the raw reconstruction, significant intensity overlap between neighboring atoms lowers tracing precision. In contrast, the GLARE-refined volume shows a marked enhancement in local spherical symmetry, with the mean value increasing from 0.50 (red box, Fig 2h-j) to 0.79 (blue box, Fig 2k-m) after optimization. GLARE refinement further reduces systematic biases in structural analyses (e.g., bond length analysis; Supplementary Fig. 7d) that originate from inaccurate atom tracing.

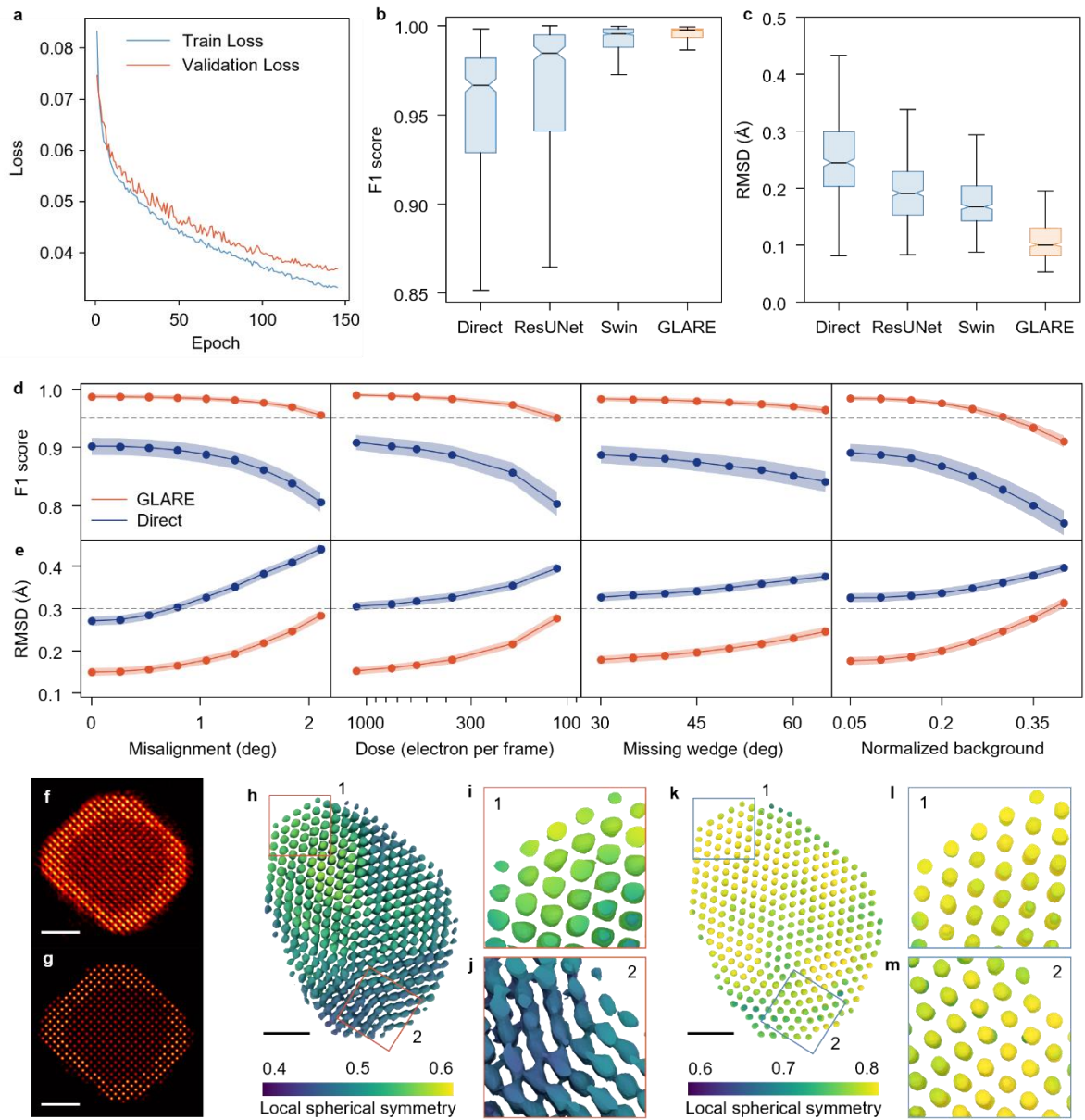

**Fig. 2 | Performance of GLARE network. a,** Training (blue) and validation (red) loss curves during 150 epochs. **b,c,** Reconstruction metrics (F1 score (**b**) and RMSD (**c**)) comparison between GLARE and other methods, including direct tracing, ResUNet, and Swin. The notches represent the 95% confidence interval regions. **d,e,** Variation of the mean F1 score (**d**) and RMSD (**e**) with different error factors, including misalignment, electron dose, missing wedge, and background intensity (from left to right respectively). The lines represent the mean value and the shaded regions represent the 99.9% confidence intervals. The red lines and shaded regions show the metrics for the atoms traced from GLARE-refined volume, while the blue lines and shaded regions show the metrics for the atoms directly traced from raw reconstructed volume. The dashed line shows the mean F1 score (**d**) and mean RMSD (**e**) on the typical-noise test set. **f,g,** The 1.37-Å-thick cross-sectional slices of the raw reconstructed volume (**f**) and the GLARE-refined volume (**g**) of an experimental Pd@Pt core-shell single-crystalline nanoparticle. Scale bars, 2 nm. **h,** An isosurface illustration of the raw reconstructed volume of an experimental Pd@Pt core-shell polycrystalline nanoparticle. The isosurface is colored according to the local spherical symmetry. Scale bar, 1 nm. **i,j,** The red boxes in (**h**) highlight Region 1 and Region 2, and magnify the region in (**i**) and (**j**), respectively. **k,** An isosurface illustration of the GLARE-refined volume of the same Pd@Pt nanoparticle shown in (**h**). The isosurface is colored according to the local spherical symmetry. Scale bar, 1 nm. **l,m,** The blue boxes in (**k**) highlight Region 1 and Region 2, and magnify the region in (**l**) and (**m**), respectively.

**Performance of the DAST network**

Figure 3 shows the performance of the DAST network designed for the element classification task. Implementation details including model architecture, dataset construction, and training process are provided in Methods. Figure 3a and b show the accuracy and the loss curves of the DAST model during 50 epochs of training, respectively. Both the training and validation losses decrease steadily, indicating stable convergence. The dashed lines represent the corresponding accuracy and loss evaluated on the test dataset. Figure 3c and Supplementary Fig. 8 compare the element classification accuracy of four methods on an identical test set: K-means method, the multilayer perceptron model (MLP), and a DAST model using only Zernike descriptors (Zern-only DAST, Supplementary Fig. 9), and the full DAST model. All neural models were trained on the same dataset. The full DAST model achieves outstanding performance with a mean accuracy of 99.50%, significantly surpassing the other methods. We also evaluated the DAST model that used only atomic coordinates as input, without the information of Zernike descriptors (Supplementary Fig. 10). After 50 epochs of training, only the training loss decreased and the training accuracy improved, whereas validation and test accuracy showed negligible improvement, plateauing around 55% (Supplementary Fig. 10). This result indicates that the atomic coordinates alone provide little useful information for the element classification task; therefore, incorporating 3D Zernike polynomial expansion coefficients as additional input features is critical to achieving high classification performance.

On the high-noise test set (Methods), DAST and K-means clustering are evaluated by classification accuracy (fraction of correctly assigned atoms). Accuracy decreases monotonically with increasing noise factors, reflecting greater uncertainty in element classification (Fig. 3d). Throughout the tested high-noise regime, DAST degrades more slowly than K-means and exhibits a narrower 99.9% confidence interval, indicating higher robustness and lower sensitivity to stochastic acquisition errors. Notably, DAST achieves a higher mean accuracy across the entire high-noise range compared with the typical-noise K-means baseline (mean accuracy = 94.16%). These results demonstrate that DAST delivers robust element classification even in challenging, dose-limited regimes.

Figure 3e-g compares the element classification performance of DAST against K-means method on a representative PdNi nanoparticle, which consist mostly Ni with a small fraction of randomly distributed Pd atoms. We compute the simulated tilt series of this nanoparticle using the multi-slice method (Methods), which ensures quantitative comparability between the simulated and experimental intensities. Figure 3e-g compares element classification on a PdNi nanoparticle. DAST (Fig. 3f) yields far fewer red-highlighted misclassifications (120 atoms) than K-means (1214 atoms; Fig. 3e) compared with the ground truth (Fig. 3g), demonstrating its superior accuracy.

We selected an experimental Pd@Pt core-shell nanoparticle to study how atomic position and intensity jointly influence element classification (Fig. 3h-j). For each atom in the nanoparticle, the local atomic intensity was quantified by integrating voxel intensities within a 3-pixel-radius sphere centered on its coordinate. Figure 3h plots this integrated intensity against the atom's distance from the nanoparticle center, revealing a correlation that aligns with the observed clustering results. Since the K-means classification method based solely on intensity, its classification is independent of an atom's distance from the surface (Fig. 3j). In contrast, the DAST model incorporates atomic coordinates, allowing it to learn a decision boundary that effectively accounts for spatial relationships such as atom-to-surface distance, thereby minimizing inter-class differences (Fig. 3i). DAST therefore achieves higher classification accuracy for atoms whose intensities fall within the intermediate range of the observed distribution, where the K-means method suffers from the strongest class overlap (Supplementary Fig. 10).

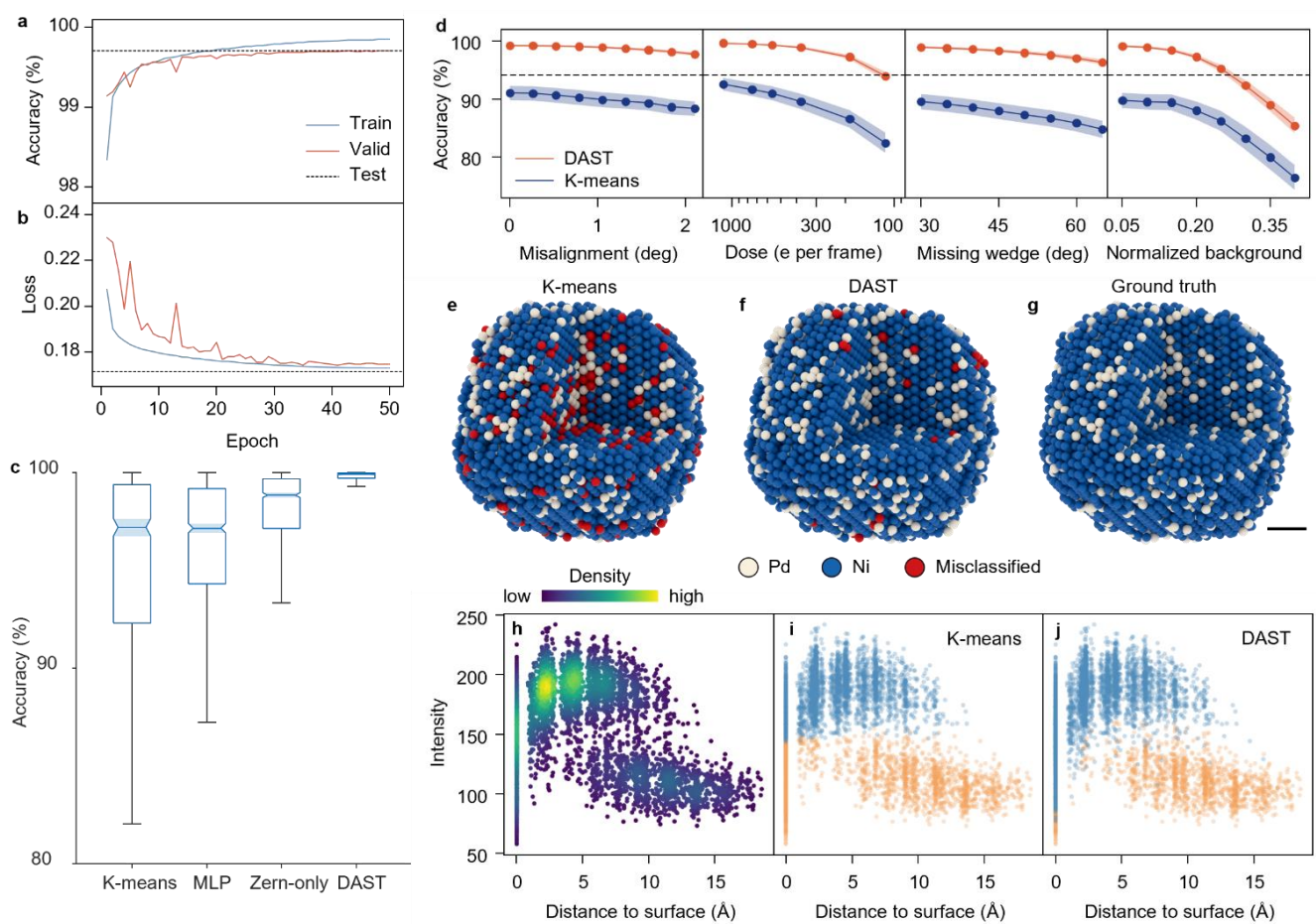

**Fig. 3 | Performance of the DAST network. a,b,** The training (blue) and validation (red) accuracy (**a**) and loss (**b**) curves during the 50 epochs. The dashed line shows the accuracy (**a**) and loss (**b**) value of the test datasets. **c**, Classification accuracy comparison between DAST and other methods, including K-means, MLP, and Zern-only DAST. The notches represent the 95% confidence interval regions. **d**, Variation of accuracy with different error factors, including misalignment, electron dose, missing wedge, and background intensity (from left to right, respectively). The lines represent the mean value and the shaded regions represent the 99.9% confidence intervals. The red lines and shaded regions show the accuracy for the performance of DAST, while the blue lines and shaded regions show the performance of K-means. The dashed line shows the mean accuracy (94.16%) of K-means on the typical-noise test set. **e-g**, The element classification performance of DAST (**f**) and K-means (**e**) on a representative PdNi nanoparticle with a ground truth atomic structure (**g**). The

nanoparticle is shown with one corner removed to reveal its internal atomic structure. The orange atoms are Pd and the blue atoms are Ni. The red atoms in (**e**) and (**f**) represent the misclassified atoms. Scale bar, 1 nm. **h-j**, The relationship between atom-to-surface distance and intensity on a Pd@Pt core-shell nanoparticle. **h**, The distribution of the atomic intensity against the atom-to-surface distance. The continuous gradient colors from purple to yellow represent the local point density. **i,j**, The K-means (**i**) and DAST (**j**) classification results. The blue points represent the Pt atoms, while the orange points represent the Pd atoms.

**Application of PANN workflow for low-dose AET data**

We further evaluated the integrated PANN workflow on experimental, dose-limited AET datasets. By combining GLARE for volume refinement and DAST for element classification, PANN enables high-performance atom tracing and element classification under low-dose conditions. Figure 4a-d and Supplementary Fig. 11 show representative 0.34-Å-thick slices of the reconstructed volumes under different conditions. Compared to the full-dose reconstruction ($5.6 \times 10^5$ e$^-$ Å$^{-2}$, Fig. 4a), reducing the dose–either by using single frame (1/3 dose) or by additionally sub-sampling the tilt series (1/6 dose; Methods)– substantially increases noise and artifacts while degrading atomic contrast in the volume (Fig. 4b; Supplementary Fig. 12a). GLARE refinement markedly enhances the clarity of atomic features in both full- and low-dose reconstructions (Fig. 4c,d; Supplementary Fig. 12b), thereby improving the interpretability of the dose-limited volumes for subsequent analysis.

We then performed atomic tracing and element classification on volumes before and after GLARE refinement to obtain atomic models for structural analysis (Fig. 4e,f; Supplementary Fig. 13). For crystalline nanoparticles, sharper and higher PDF peaks signify greater atomic order and reduced lattice distortion, aligning with the intrinsic physical properties of crystalline systems. Our PDF analysis reveals that peaks progressively sharpen and intensify from the lowest dose baseline to PANN-derived structure (Fig. 4e; Supplementary Fig. 13a), indicating reduced lattice distortion and improved recovery of crystalline order. Notably, even at 1/6 of the full dose, the PANN-derived structure produces sharper PDF peaks than the full-dose baseline workflow (Supplementary Fig. 13a). A consistent trend is observed in the bond-length statistics. The bond-length distributions systematically narrow for PANN across all dose levels (Fig. 4f; Supplementary Fig. 13b), indicating more uniform local geometry and a physically more plausible structural model. This increased uniformity corresponds to a more homogeneous energy landscape, which is a physically more rational outcome.

Figure 4g,h presents the consistency and RMSD correlation analysis of the atomic structures among different dose conditions. We quantified robustness using cross-condition consistency and repeatability analyses by comparing structures from independent reconstructions (Fig. 4g,h, Methods). Across dose conditions, atomic structures generated by PANN exhibit higher mutual consistency than those from the baseline workflow (Fig. 4i). In the most dose-limited case (1/6 dose), two independent PANN-refined

reconstructions (P-1/6) show 95% agreement, whereas the corresponding structures from the conventional direct-reconstruction method (C-1/6) only reach 75% consistency. This improvement is further reflected in the RMSD metrics. The RMSD between two independent P-1/6 structures is 0.3 Å, substantially lower than the 0.5 Å for C-1/6 (Fig. 4h). For experimental AET data, where ground-truth coordinates are unavailable, correlation analysis similarly confirms that PANN-derived structures maintain higher mutual consistency, underscoring the reliability of the PANN workflow for dose-limited AET post-processing.

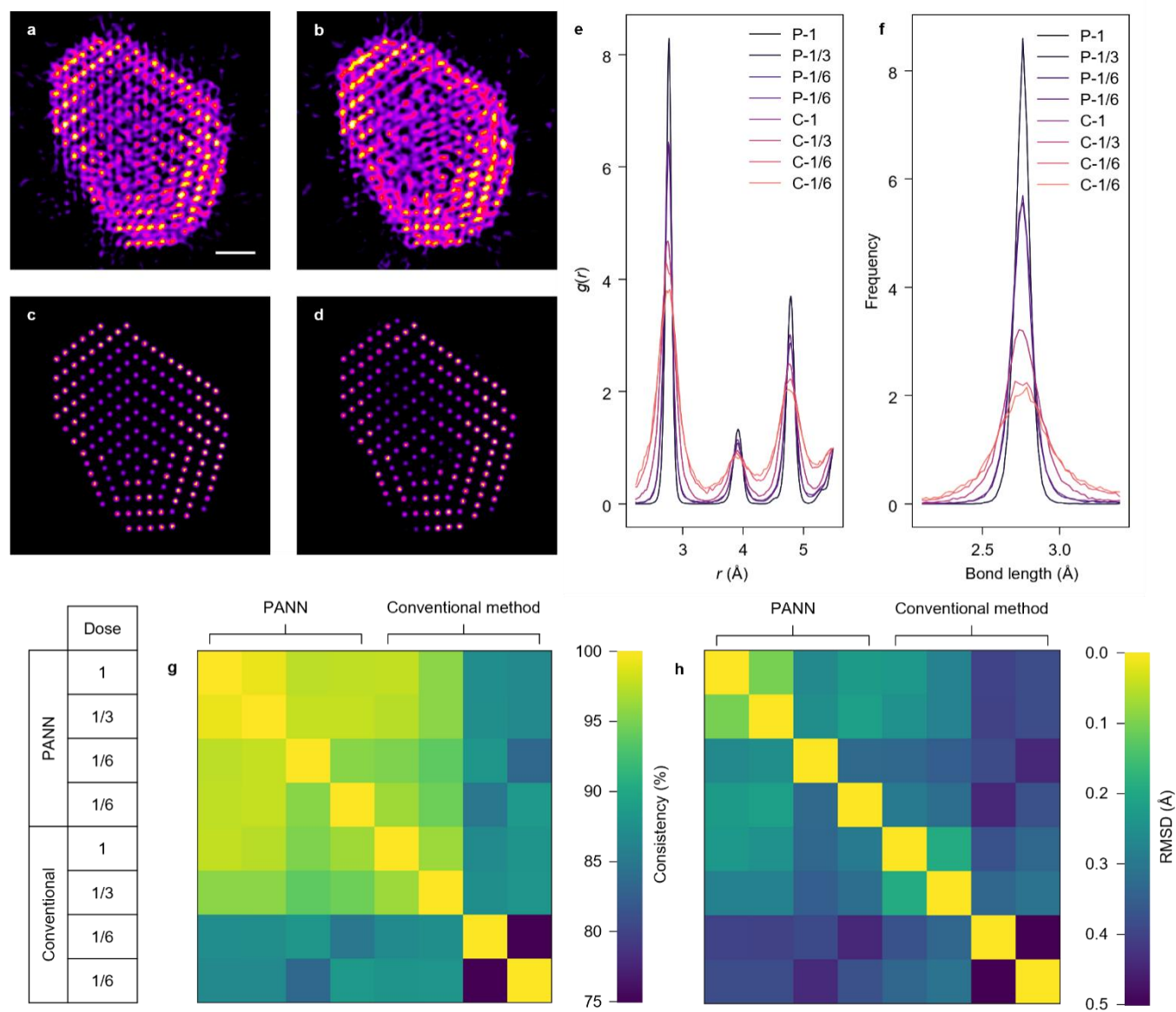

**Fig. 4 | Performance of the PANN workflow on the experimental data. a-d**, The 0.34-Å-thick slices of raw (**a,b**) and refined (**c,d**) reconstructed volume under typical-dose (**a,c**) and 1/6-dose (**b,d**) conditions. Scale bar, 1 nm. **e,f**, PDF (**e**) and bond length (**f**) analysis of the atomic structures yielded from different workflows. **g,h**, The consistency (**g**) and RMSD (**h**) correlation analysis among the atomic structures. In the correlation matrices, from top to bottom and left to right, each row and column correspond sequentially to P-1, P-1/3, P-1/6 (odd), P-1/6 (even), C-1, C-1/3, C-1/6 (odd), and C-1/6 (even). P denotes the atomic structures obtained using the PANN workflow, including GLARE refinement and DAST element classification, while C denotes the atomic structures obtained using the conventional post-processing workflow (Methods); the suffixes 1, 1/3, and 1/6 denote the typical-, 1/3-, and 1/6-dose tilt series reconstructions, respectively.

**Fine-tuning of GLARE for the model's generalization capability**

Next, we fine-tuned the GLARE neural network to extend its applicability to electron beam sensitive materials such as halide perovskites and quantum dots that were absent from the pre-training dataset. As an example, $CsPbBr_3$ is a widely studied perovskite model platform for optoelectronic applications, with nanoscale structure-property relationships closely linked to local composition, defects and lattice distortions. However, $CsPbBr_3$ is highly beam sensitive (dose tolerance $\leq$ 8,000 e$^-$ Å$^{-2}$)[49], rendering conventional-dose HAADF-STEM AET tilt-series acquisition impractical. We simulated low-dose (5,000 e$^-$ Å$^{-2}$) HAADF-STEM tilt series of a $CsPbBr_3$ nanoparticle using multi-slice simulation (Fig. 5a,b, Supplementary Fig. 14) and reconstructed the corresponding volumes.

To improve the reconstruction quality, we applied GLARE to the low-dose AET data. Since this material class was absent from the original training distribution, direct application of the pre-trained model introduced a domain shift, resulting in ghost-atom artifacts within the refined volumes (Supplementary Fig. 15). A 0.34-Å-thick slice of the reconstructed volume shows pronounced noise (Fig. 5c,d). Direct atom tracing on this volume achieved low consistency with the ground-truth atomic model, yielding an F1 score of 0.8058 and an RMSD of 0.3484 Å, indicating that the original structure could not be reliably recovered under these low-dose conditions. To adapt the pre-trained GLARE model to $CsPbBr_3$ without full retraining, we constructed a small simulated fine-tuning dataset (300 volumes). This yielded a fine-tuned specialized model, termed GLARE-CPB. Atom tracing on the volume refined by GLARE-CPB (Fig. 5e,f) yielded substantially improved atomic coordinates, with an F1 score of 0.9715 and an RMSD of 0.1269 Å. This performance enables accurate determination of almost all the atoms within the nanoparticle. Then we evaluated how fine-tuning dataset size influences performance. As the number of fine-tuning samples was reduced from 300 to 200, 100, 50, and 25, atom-tracing performance progressively degraded, evidenced by a drop in F1 score and an increase in RMSD (Fig. 5g, h). Notably, with only 200 training samples, atom-tracing accuracy already reached a practically applicable level (F1 = 0.9696, RMSD = 0.1296 Å), indicating that reliable material-specific adaptation can be achieved with a relatively limited task-specific data. Therefore, A fine-tuning set of ~200 samples offers a favorable balance between annotation effort and reconstruction accuracy. We similarly fine-tuned GLARE for other materials including ZSM-5 (Supplementary Figs. 16,17) and CdSe quantum dot (Supplementary Figs. 18,19) with 200 fine-tuning samples, achieving strong performance (Supplementary Tables 3 & 4). These results highlight our model demonstrates strong generalization capability, enabling robust refinement across diverse material systems and experimental conditions.

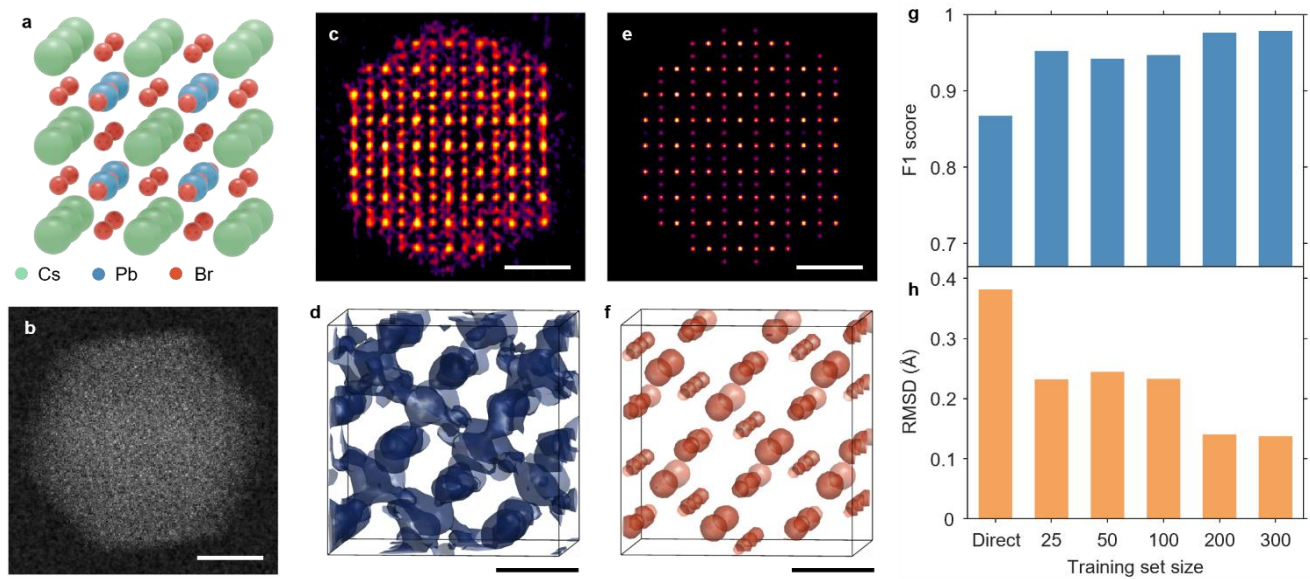

**Fig. 5 | Fine-tuning of GLARE for CsPbBr₃ reconstruction. a**, The atomic model of crystalline $CsPbBr_3$ for the following test. **b**, Multi-slice simulated 0° HAADF-STEM image of the $CsPbBr_3$ nanoparticle under low-dose conditions. Scale bar, 2 nm. **c**, A 0.34-Å-thick slice from the reconstructed volume obtained from the low-dose tilt series. Scale bar, 2 nm. **d**, Representative isosurface rendering of a region extracted from the raw reconstructed volume. Scale bar, 5 Å. **e**, Corresponding slice to that in **c** from the volume refined by GLARE-CPB model. Scale bar, 2 nm. **f**, Isosurface rendering of the corresponding region in **d** from the volume refined by the GLARE-CPB model. Scale bar, 5 Å. **g,h**, Performance of the fine-tuned GLARE-CPB model trained on datasets of different sizes, evaluated by F1 score (**g**) and RMSD (**h**).

## Discussion

In this work, we introduce PANN an integrated, physics-aware neural network workflow for efficient, robust, generalizable and accurate determination of 3D atomic structure determination from low-dose AET data. PANN combines a global-local 3D ResUNet (GLARE) to refine the reconstructed volume for precise atom tracing, and subsequently a graph attention set Transformer (DAST) to determine the element species of each atom. Evaluated extensively on simulated benchmarks and experimental data, PANN demonstrates superior performance under high-noise conditions, significantly improving both atomic positional precision and elemental accuracy over conventional methods. Crucially, the workflow enables high-fidelity reconstruction at substantially reduced electron doses, mitigating beam-induced damage while maintaining structural integrity. These advances highlight PANN's generalizability and its potential to expand the frontiers of atomic-scale tomography for beam-sensitive nanomaterials.

Despite these advancements, the current PANN framework has several limitations that point to future directions. First, it assumes the imaged nanostructure remains static during tilt-series acquisition. In practical scenarios where beam-induced or intrinsic dynamics occur, this static reconstruction assumption may limit reliability. We anticipate that training GLARE on temporally resolved or non-static AET data would enable its application to dynamically evolving systems. Second, PANN operates as a modular post-processing pipeline, depending on specific upstream data preprocessing. Developing a fully end-to-end trained workflow–from raw tilt series to final atomic model–would improve overall consistency and integration. Finally, DAST is currently validated only for binary elemental systems. Expanding its training to ternary or multi-component materials, and to datasets that explicitly include ghost-atom artifacts, would broaden its applicability to more complex material chemistries.

## Methods

### Dataset generation

All training data were generated through a simulation pipeline comprising six stages: (i) nanoparticle generation, (ii) creation of a ground-truth target volume, (iii) sampling of tilt-series projection orientations, (iv) simulation of HAADF-STEM tilt series, (v) injection of noise and experimental imperfections, and (vi) 3D reconstruction to produce the network input volume. Nanoparticle structures were built using the atomic simulation environment (ASE) package in Python, whereas all subsequent simulation and reconstruction steps were implemented in MATLAB.

### Nanoparticle generation and structural relaxation

Initial nanoparticle structures were constructed with randomized key attributes—including elemental composition/stoichiometry, size, structural category (single-crystalline, polycrystalline, or amorphous), morphology, and phase (e.g., solid solution or intermetallic compound)—to ensure structural diversity. These coarse configurations were then physically relaxed to improve realism. For crystalline (single- and poly-crystalline) particles, molecular statics with an embedded-atom method (EAM) potential was used. For amorphous particles, molecular dynamics was employed: atoms were randomly placed within a sphere (subject to a minimum interatomic separation), heated to a high temperature (1000–2000 K), and rapidly quenched to a low temperature (100–300 K). Configurations sampled along this thermal cycle yielded a diverse set of disordered states, from gas-like and liquid-like to glassy structures.

### Construction of the ground-truth (standard) volume

The ground-truth volume was constructed directly from atomic coordinates by convolving each atomic position with an isotropic 3D Gaussian kernel, centered at the atom. The standard deviation of the Gaussian was set as

$$\sigma = \frac{1}{\pi s}\sqrt{\frac{B}{8}} \qquad (1)$$

where $B$ is the Debye–Waller B-factor (fixed to 12 Å$^2$ in our simulations) and $s$ denotes the voxel size[50]. To approximate HAADF Z-contrast, the peak amplitude of each Gaussian was scaled proportionally to $Z^{1.7}$, where $Z$ is the atomic number[51]. Finally, the generated volume was normalized to the range [0, 1].

**Construction of HAADF-STEM tilt series**

After generating the ground-truth volume, a set of tilt angles was defined. HAADF-STEM projections were generated via forward projection,

$$I_{\text{HAADF}} = \Pi_\theta(\rho_R O\{x,y,z\}) = (\Pi_\theta O\{u,v,w\}) = \int O\{u,v,w\}dw \qquad (2)$$

where $I_{\text{HAADF}}$ is the simulated HAADF-STEM image, $\Pi_\theta$ denotes the projections operator at tilt angle $\theta$, $\rho_R$ rotates the volume by $\theta$ about the rotation axis, $O$ is the 3D volume, and $(u, v, w)$ are coordinates in the rotated frame, defined by

$$(u,v,w)^T = R^{-1}(x,y,z)^T \qquad (3)$$

where $R$ is the corresponding rotation matrix.

To emulate realistic AET acquisition conditions, the following parameters were randomly sampled within predefined ranges and applied during simulation:

(1) Voxel size: 0.343–0.352 Å;
(2) Angular range: 135°–155°, with goniometer angle errors modeled as Euler-angle perturbations up to (2°, 2°, 2°);
(3) Translational alignment errors: modeled as a Gaussian distribution, $\sigma = 0.2$ pixel;
(4) Electron dose per projection: $10^3$–$10^4$ e·Å$^{-2}$;
(5) Additive Gaussian image noise: $\sigma = 1\%$;
(6) Substrate (background) contrast: 5%–15%;
(7) Defocus: modeled by a 2D Gaussian point-spread function with $\sigma = 1$–1.5 pixels.

These parameters were used both in the forward-projection step and in the subsequent 3D reconstruction, producing paired input (reconstructed) and target (ground-truth) volumes.

To approximate the spatially correlated background fluctuations ($B(\mathbf{r})$) typical of experimental HAADF-STEM images, a correlated random background field was added instead of pixel-independent noise[52]. Specifically, an initially uncorrelated 2D random field $\xi(\mathbf{r})$ was convolved with a normalized Gaussian kernel $G_\sigma(\mathbf{r})$:

$$B(\mathbf{r}) = (G_\sigma * \xi)(\mathbf{r}) \qquad (4)$$

where $\sigma$ defines the correlation width (set to 1 pixel, $\approx 0.35$ Å, unless noted otherwise). This suppresses high-frequency components and introduces a finite lateral correlation length, mimicking the slowly varying background. The resulting background field was normalized and added to the simulated images at a prescribed amplitude.

**Reconstruction and atom tracing (Conventional processing workflow)**

Following the generation of the simulated tilt series, projections were preprocessed using a standard pipeline including denoising and background subtraction. BM3D was used for image denoising[53]. For each denoised projection, the background was estimated using the discrete Laplacian function in MATLAB and subsequently subtracted. Finally, the total intensity of each projection was normalized so that all images in the tilt series had the same integrated intensity. After preprocessing, 3D reconstruction was performed using the RESIRE algorithm to obtain the reconstructed volume[21].

Atom tracing and element classification were performed on the reconstructed volume. The reconstructed volume was first up-sampled using spline interpolation. All local maxima in the interpolated volume were identified as candidate atomic positions. To filter out low-intensity non-atomic peaks, the integrated intensity within a 3×3×3 voxel region around each candidate was computed, and K-means clustering applied to these sums removed the low-intensity cluster. A minimum-distance constraint of 2 Å was then enforced to eliminate overly close positions. Finally, chemical species were assigned by performing K-means clustering on the integrated intensity within a 2.4×2.4×2.4 Å$^3$ (7×7×7 voxel) region centered on each remaining candidate, using the originally generated atomic model as the reference.

**3D Zernike expansion**

For each localized atom, a 3D Zernike expansion was performed to obtain the corresponding expansion coefficients[54,55]. The real-valued, normalized generalized 3D Zernike expansion on the unit ball is defined as

$$f(\mathbf{r}) = \sum_{n=0}^{\infty}\sum_{l=0}^{\infty}\sum_{m=-l}^{l} \alpha_{nlm} Z_{nlm}(\mathbf{r}), \qquad \mathbf{r} \in \mathbb{B}^3 = \{\mathbf{r}: \|\mathbf{r}\| \leq 1\} \tag{5}$$

with basis functions

$$Z_{nlm}(\mathbf{r}) = \bar{R}_{nl}^{1}(r) Y_{lm}^{(\text{real})}(\theta, \phi), \qquad r = \|\mathbf{r}\| \tag{6}$$

where $Y_{lm}^{(\text{real})}$ denotes real spherical harmonics, $\bar{R}_{nl}^{p}(r)$ is the normalized 3D Zernike radial polynomial. The real spherical harmonics $Y_{lm}^{(\text{real})}$ are given by

$$Y_{lm}^{(\text{real})}(\theta, \phi) = (-1)^m \sqrt{\frac{2l+1}{4\pi}\frac{(l-|m|)!}{(l+|m|)!}} P_l^{|m|}(\cos\theta) \begin{cases} 1, & m = 0 \\ \sqrt{2}\cos(m\phi), & m > 0 \\ \sqrt{2}\sin(|m|\phi), & m < 0 \end{cases} \tag{7}$$

where $P_l^{|m|}$ are the associated Legendre polynomials defined by

$$P_l^m(x) = (-1)^m \frac{(1-x^2)^{m/2}}{2^l l!} \frac{d^{l+m}}{dx^{l+m}}(x^2-1)^l. \tag{8}$$

The normalized 3D Zernike radial polynomial $\bar{R}_{nl}^{p}(r)$ is expressed as

$$\bar{R}_{nl}^{p}(r) = \mathcal{N}_{nl}^{p} R_{nl}^{p}(r)$$

$$\mathcal{N}_{nl}^{p} = \sqrt{4n + 2l + p + 2}$$

$$R_{nl}^{p}(r) = (-1)^n x^l P_n^{[(l+p/2),0]}(1-2x^2) \tag{9}$$

where $P_n^{(\alpha,\beta)}$ are the Jacobi polynomials defined by

$$P_n^{(\alpha,\beta)}(x) = \frac{(-1)^n}{2^n n!}(1-x)^{-\alpha}(1+x)^{-\beta}\frac{d^n}{dx^n}[(1-x)^\alpha(1+x)^\beta(1-x^2)^n]. \tag{10}$$

Central slices of selected basis functions are shown in Supplementary Fig. 20.

**Calculation of local spherical symmetry**

The 3D Zernike polynomials decomposed the local atomic volume into components with different symmetries. As shown in Supplementary Fig. 19, terms with $l = 0$ corresponded to spherically symmetric contributions, while all terms with $l > 0$ captured non-spherical components. The fractional contribution of the $l = 0$ coefficients provided a quantitative measure of the local spherical symmetry (LSS) around an atom,

$$\text{LSS} = \frac{\sum_{n,m} \alpha_{n0m}}{\sum_{n,l,m} \alpha_{nlm}}. \tag{11}$$

**Multi-slice image simulation**

Unless otherwise noted, all multi-slice simulations were performed using Prismatic 2.0 implemented in Python[56-58]. For each simulated projection, the atomic model at 0° was rotated according to the Euler-angle series using an extrinsic rotation convention. HAADF-STEM images were calculated with 4 frozen-phonon configurations under the following conditions: accelerating voltage 300 kV, a probe semi-angle 30 mrad, pixel size 0.3434 Å, and an annular detector collection range 39.4–200 mrad. The probe was focused at the center of the nanoparticle along the beam direction (z axis). After image formation, Poisson noise corresponding to the collected electron count (determined from the screen current) was added, and the simulated images were further convolved with a point-spread function with $\sigma = 1\text{–}1.5$ pixels.

**GLARE implementation**

All models were developed in PyTorch (v2.8.0) with CUDA (v12.6.0) and trained on a cluster using four NVIDIA A100 GPUs (80 GB memory each). Distributed training was performed with PyTorch's DistributedDataParallel (DDP) over the NCCL backend, initialized from environment variables.

**Model construction**

The GLARE network was built based on a 3D residual U-Net with a global-local feature modulation branch. The encoder consisted of four convolutional stages, with the channels width doubling at each stage starting from 16 channels in the first stage. Each encoder stage contained a 3×3×3 convolution, followed by normalization, a leaky-ReLU activation (slope = 0.1), and two residual blocks. Down-sampling between adjacent stages was performed using 2×2×2 max-pooling.

Each residual units contained two 3×3×3 convolutions with normalization. The first convolution was followed by an activation, and the residual summation was passed through a leaky-ReLU layer. Activations were implemented by a leaky-ReLU layer with negative slope 0.1, and normalization used 3D group normalization layer.

At the bottleneck, a 3×3×3 convolution with normalization and activation was applied. Global context was extracted from the bottleneck tensor using adaptive average pooling to 1×1×1 vector, which was then passed through two fully connected layers (with a ReLU in between) to produce a fixed-length global embedding of 128 dimensions.

The decoder up-sampled features using 2×2×2 transposed convolution. At each level, the up-sampled feature was concatenated with the corresponding skip connection from the encoder, followed by a post-fusion block consisting of convolution, normalization, activation, and residual units. To inject global information, a feature-wise linear modulation (FiLM) layer was applied at every decoder levels. For a decoder feature map $x$ and global embedding $g$, FiLM computed per-channel scaling $\gamma(g)$ and shifting $\beta(g)$ via linear projections, and modulates the feature as: $x \leftarrow x \cdot (1 + \gamma) + \beta$. The network finally output a single-channel denoised volume.

**Training**

The model was trained on paired 3D volumes stored in MATLAB format (.mat), with one volume serving as the input (reconstructed tomogram) and the corresponding volume as the target (ground-truth). All volumes were cropped or padded to a fixed size of 256×256×256 voxels and independently standardized (z-score normalization). After preprocessing, each volume was converted to a tensor of shape [1, D, H, W].

The full dataset was split into training and validation sets, with 10% of the data reserved for validation. For multi-GPU training, distributed samplers were used for both subsets. Training ran for 200 epochs using the AdamW optimizer (learning rate $1\times10^{-4}$, weight decay $1\times10^{-4}$) with a cosine-annealing learning-rate schedule over the entire training span. Mixed-precision training was enabled via PyTorch's Automatic Mixed Precision (AMP) with gradient scaling. The training objective was voxel-wise regression from the reconstructed volume to the target volume, optimized with an $L_1$ loss. Validation was performed at the end of every epoch.

**Evaluation (computation of F1 score and RMSD on the test set)**

Model performance was evaluated on a held-out test set (subset fraction = 0.1) consisting of 507 samples. For each sample, atomic coordinates were obtained in two ways: (1) Direct tracing from the original reconstructed volume, yielding coordinates ($p_D$). (2) GLARE-refined tracing from the volume processed by the trained GLARE model, yielding coordinates ($p_G$).

To compare both sets of predicted coordinates ($p_D$ and $p_G$) with the ground-truth coordinates ($p_T$), all coordinate sets were aligned using the iterative closest point algorithm[59]. After alignment, the matched predicted atoms were denoted as $\tilde{p}_D$ and $\tilde{p}_G$, and their corresponding ground-truth matches as $\tilde{p}_{D,T}$ and $\tilde{p}_{G,T}$, respectively. Taking direct tracing as an example, precision, recall, and the F1 score were computed as

$$\text{Precision} = \frac{\text{TP}_D}{\text{TP}_D + \text{FP}_D}$$
$$\text{Recall} = \frac{\text{TP}_D}{\text{TP}_D + \text{FN}_D}$$
$$\text{F1} = \frac{2 \cdot \text{Precision} \cdot \text{Recall}}{\text{Precision} + \text{Recall}}. \tag{12}$$

where $\text{TP}_D$ (true positive) is the number of correctly matched atoms in $p_D$; $\text{FP}_D$ (false positive) is the number of ghost atoms present in $p_D$ that have no match in $p_T$. $FN_D$ (false negative) is the number of ground-truth atoms missing from $p_D$. These counts were obtained as:

$$TP_D = N(\tilde{p}_D), \quad FP_D = N(p_D) - TP_D, \quad FN_D = N(p_T) - TP_D \tag{13}$$

where $N(\cdot)$ denotes the number of atoms in a coordinate set.

RMSD for the directly traced atoms was computed as:

$$RMSD = \sqrt{\frac{\sum_{i=1}^{N(\tilde{p}_D)} \left\| \tilde{p}_D^{(i)} - \tilde{p}_{D,T}^{(i)} \right\|_2^2}{N(\tilde{p}_D)}}. \tag{14}$$

The same metrics were calculated analogously for the GLARE-refined coordinates $p_G$.

## DAST implementation

### Dataset preprocessing

DAST was trained using MATLAB data files generated for individual nanoparticles. In each file, the input was stored as an atom-wise feature matrix with 228 columns, together with a corresponding one-dimensional label vector. For each atom, the first three columns encoded the Cartesian coordinates $(x, y, z)$, and the remaining 225 columns contained precomputed 3D Zernike expansion coefficients serving as structural descriptors.

To reduce scale variability across nanoparticles, atomic coordinates were centered and normalized by the root-mean-square (RMS) distance within each nanoparticle. Per-atom descriptor vectors were independently standardized per nanoparticle using a feature-wise zero-mean, unit-variance transform (with numerical safeguards for near-zero variances). Both operations were applied on-the-fly during data loading. All MATLAB files in a specified root directory were collected and deterministically split into training, validation, and test subsets via a seeded random shuffle, with a default split ratio of 0.8 :0.1 :0.1. Because nanoparticles contain a variable number of atoms, samples within a batch were padded to the maximum atom count in that batch. A boolean mask identified valid (non-padded) atom entries, and labels for padded positions were set to -100 so that they are ignored by the cross-entropy loss.

### Model construction

DAST was designed to classify the element of each atom in a variable-sized nanoparticle. The default architecture used a hidden dimension of 160, 4 attention blocks, 6 attention heads, a 12-nearest-neighbor graph, and 3 output classes.

For each atom, the 225-dimensional descriptor vector was first normalized using LayerNorm and then linearly projected to the hidden dimension. The atomic coordinates were separately projected to the same hidden dimension, and the two embeddings were summed to produce the initial token representation.

For each nanoparticle, a k-nearest-neighbor (k-NN) graph was constructed on the basis of pairwise Euclidean distances computed using the torch.cdist function. Distances involving padded atoms were assigned a large value to exclude invalid neighbors. For each atom, the K smallest distances were selected, and a corresponding neighbor-validity mask was generated. When the number of atoms was smaller than K, the neighbor lists were padded and the added entries were masked accordingly. Neighbor distances were encoded using a radial basis function (RBF) expansion, with centers linearly spaced between 0 and 6. A learnable scale parameter $\gamma$ controlled the width of the basis functions.

Each attention block applied a neighbor-restricted multi-head attention operator, in which each atom attended only to its k-NN set. An edge bias was computed by concatenating (i) the RBF distance features and (ii) the relative displacement vector between the atom and its neighbor, and then passing the result through a small MLP to produce a scalar bias for each edge, which was broadcast across attention heads. Invalid neighbors were masked before softmax normalization. Residual connections were applied around the neighbor attention module and around a position-wise feed-forward network (MLP) with dropout. After the final attention block, per-atom logits were obtained by applying LayerNorm followed by a linear classifier projecting to the three element classes.

**Training and evaluation**

The model was trained by minimizing the masked cross-entropy loss, where padded atom positions were ignored. Optimization was performed with the AdamW optimizer (learning rate $3\times10^{-4}$, weight decay $1\times10^{-2}$) and a cosine-annealing learning-rate schedule over 50 epochs by default. Training employed AMP and gradient scaling on CUDA devices.

Masked accuracy was computed by comparing the argmax predictions with ground-truth labels only over valid (non-padded) atoms. The best model checkpoint was selected based on validation accuracy and saved for final evaluation.

**Experimental low-dose consistency test for AET tilt series**

To assess reconstruction consistency under low-dose conditions, reduced-dose datasets were generated from an experimentally acquired full-dose tilt series from literature[26]. In the original acquisition, three consecutive HAADF-STEM images were recorded at each tilt angle and summed to produce a high-signal-to-noise projection for the standard reconstruction.

To simulate a 1/3-dose condition, the three sequential images at each angle were treated as independent projections, each representing approximately 1/3 of the total dose per tilt. A 1/6-dose dataset was created by first using only single-frame images (as in the 1/3-dose case) and further dividing the tilt series into two independent subsets based on acquisition order—retaining only the odd-numbered or even-numbered projections. This reduced both the number of projections and the dose per projection, yielding an effective dose of about 1/6 of the original.

Notably, the 1/6-dose condition exhibits a significantly enlarged missing wedge due to the reduced projection count; thus, it does not represent an optimally sampled low-dose acquisition, and better performance could likely be achieved with a more optimized tilt-series design.

Both the 1/3-dose and 1/6-dose datasets were independently reconstructed using the same conventional AET workflow applied to the full-dose data. Consistency between two atomic models was quantified as the number of atom pairs matched by iterative closest point within a 1 Å distance threshold, divided by the average total number of atoms in the two models.

**Acknowledgments:**

This work was supported by the National Natural Science Foundation of China (Grant No. 92477203 & 22172003), the National Key R&D Program of China (Grant No. 2024YFA1509500), Beijing Natural Science Foundation (Grant No. F251007) and the High-performance Computing Platform of Peking University.

**Author contributions:**

J.Z. and Y.Z. conceived the idea. Y.Z. & L.C. developed the methodology, performed the simulations and data analysis, implemented the GLARE and DAST models, and drafted the manuscript under the supervision of J.Z. Y.Z., L.C. & Z.S. tested the models on experimental data. J.Z. supervised the project, contributed to the conceptual development of the study, discussed the results, wrote and revised the manuscript. All authors commented on the manuscript.

**Competing interests:**

A Chinese patent that covers the GLARE and DAST models for atom tracing and element classification reported in this paper was filed by Peking University (application 202610074547.9).

**Supplementary information:**

Supplementary Tables 1–4.

Supplementary Figs. 1–20.

# Supplementary Information for

# Physics aware neural networks enable robust and full atomic structure determination via low dose atomic electron tomography

Yao Zhang[1], Lanyi Cao[1], Zhen Sun[1], Jihan Zhou[1]*

[1]*Beijing National Laboratory for Molecular Sciences, Center for Integrated Spectroscopy, College of Chemistry and Molecular Engineering, Peking University; Beijing 100871, China*

*Correspondence and requests for materials should be addressed to J. Z. (email: jhzhou@pku.edu.cn)*

**Supplementary Table 1 | Acquisition parameters for the single crystalline Pd@Pt nanoparticle comprises ≈8800 atoms.**

| | |
|---|---|
| Voltage (kV) | 300 |
| Convergence semi-angle (mrad) | 30.0 |
| Detector inner angle (mrad) | 39.4 |
| Detector outer angle (mrad) | 200 |
| Pixel size (Å) | 0.343 |
| Number of projections | 60 |
| Tilt range (°) | -75.5~77.5 |
| Electron dose ($10^5$ e$^-$·Å$^{-2}$) | 5.7 |

**Supplementary Table 2 | Acquisition parameters for the polycrystalline Pd@Pt nanoparticle comprises ≈5400 atoms.**

| | |
|---|---|
| Voltage (kV) | 300 |
| Convergence semi-angle (mrad) | 30.0 |
| Detector inner angle (mrad) | 39.4 |
| Detector outer angle (mrad) | 200 |
| Pixel size (Å) | 0.343 |
| Number of projections | 59 |
| Tilt range (°) | -76~77 |
| Electron dose ($10^5$ e$^-$·Å$^{-2}$) | 5.6 |

**Supplementary Table 3 | F1 scores and RMSDs of the atomic coordinates traced from ZSM-5 volumes refined using fine-tuned GLARE models.**

|  | F1 score | RMSD (Å) |
|---|---|---|
| Raw reconstruction | 0.718 | 0.579 |
| Refined reconstruction | 0.958 | 0.203 |

**Supplementary Table 4 | F1 scores and RMSDs of the atomic coordinates traced from CdSe volumes refined using fine-tuned GLARE models.**

|  | F1 score | RMSD (Å) |
|---|---|---|
| Raw reconstruction | 0.763 | 0.471 |
| Refined reconstruction | 0.957 | 0.215 |

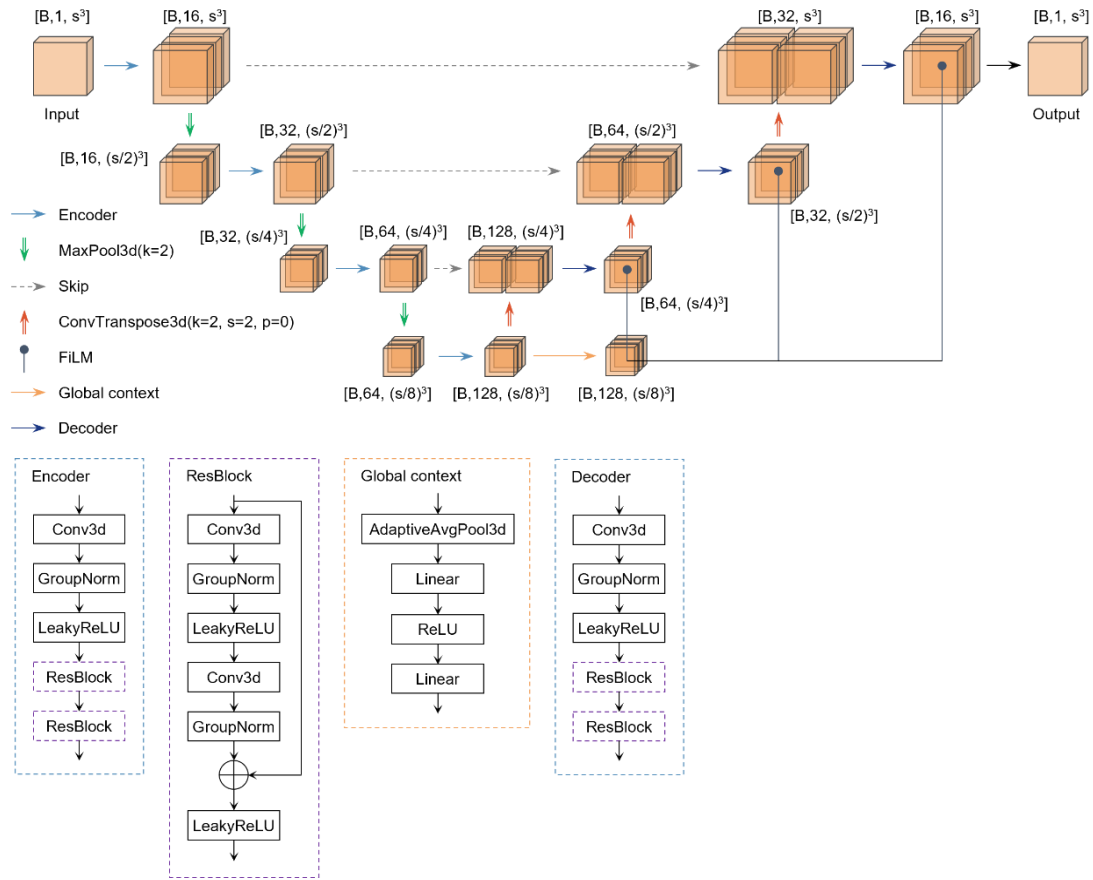

**Supplementary Figure 1 | Schematic of the GLARE network architecture.** The network accepts a 3D volume as input and is built on a four-level ResUNet backbone. In addition to the standard ResUNet design, a global-context branch is introduced to expand the effective receptive field of the network.

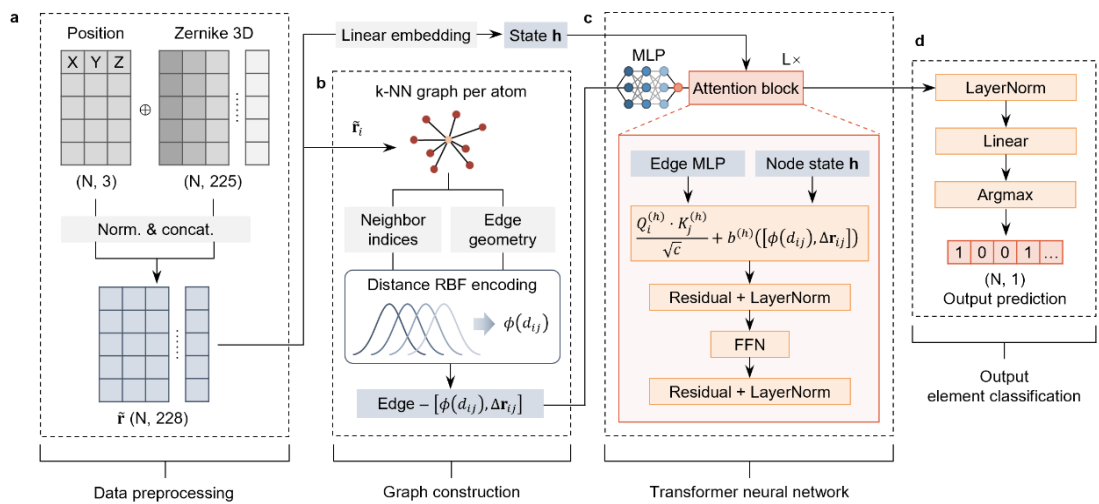

**Supplementary Figure 2 | Schematic of the DAST network architecture.** DAST accepts an $N \times 228$ feature matrix formed by concatenating normalized atomic coordinates with their corresponding 3D Zernike expansion coefficients. For each atom, a local neighborhood graph is constructed using k-nearest neighbors (k-NN), and interatomic distances are encoded via a radial basis function (RBF) expansion. The encoded edge features are processed by an MLP and combined with linearly embedded node features as input to a stack of L attention blocks. Chemical species information for each atom is incorporated at the final stage.

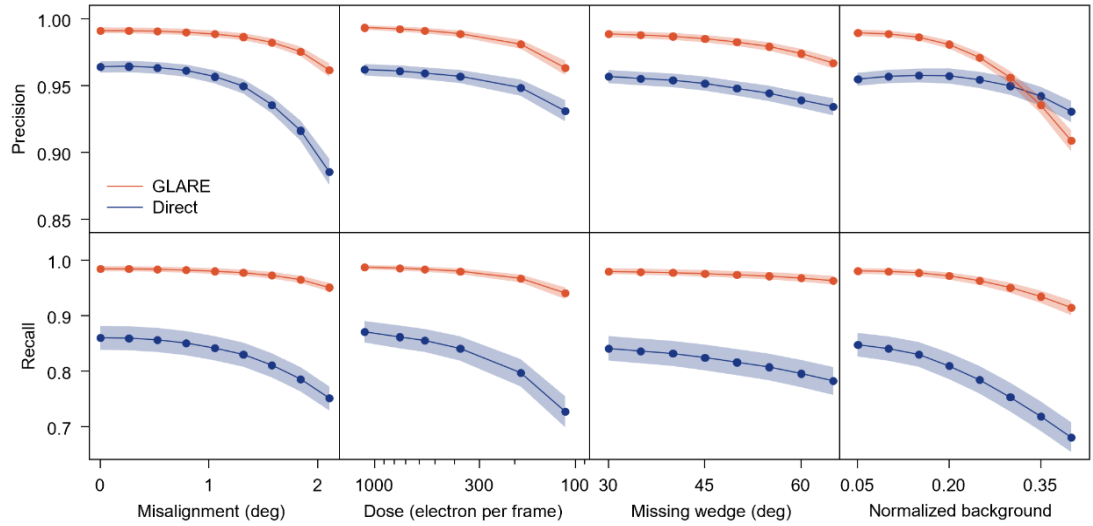

**Supplementary Figure 3 | Precision and recall of GLARE on the test set.** The red curves denote the precision and recall achieved by GLARE, while the blue curves represent the performance of atomic localization obtained by direct tracing from the reconstructed volumes.

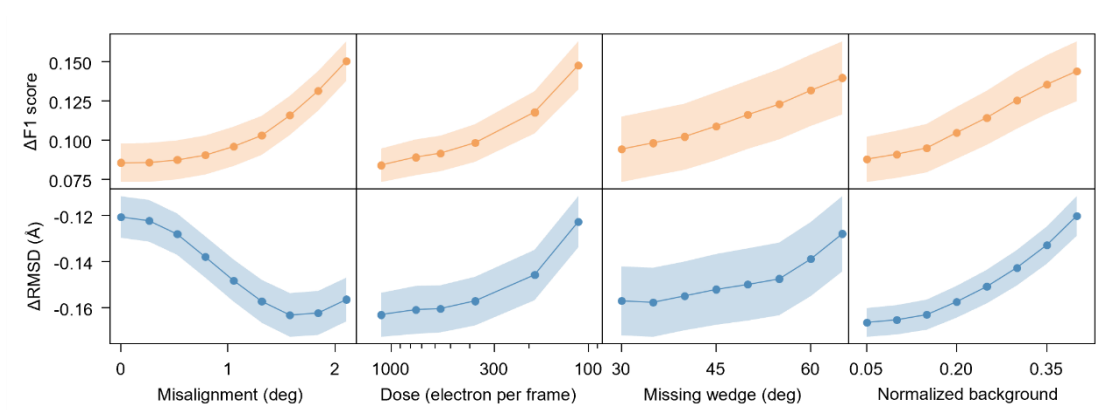

**Supplementary Figure 4 | Performance difference between GLARE and direct tracing.** Difference in F1 score and RMSD between atom tracing by GLARE and direct tracing from the reconstructed volumes on the test set (GLARE - direct).

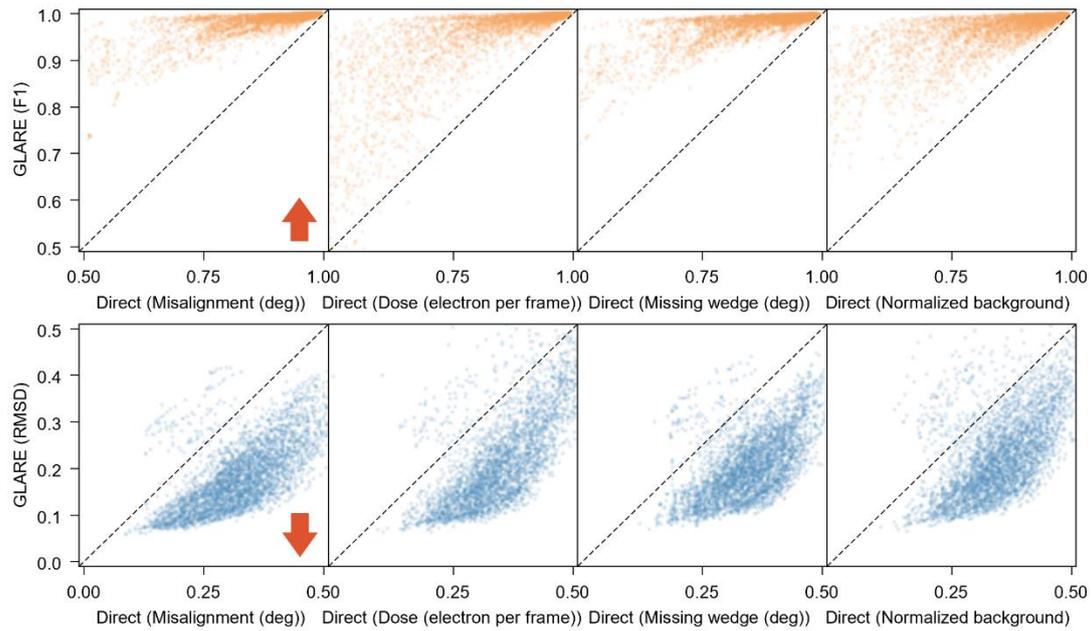

**Supplementary Figure 5 | Scatter plots comparing GLARE with direct tracing.** The *x*-axis shows results from direct atom tracing, and the y-axis shows results after GLARE refinement. The top row displays the F1 metric, where points above the diagonal indicate improved performance with GLARE. The bottom row displays the RMSD metric, where points below the diagonal correspond to improvement with GLARE. The dashed line marks the *y* = *x* reference.

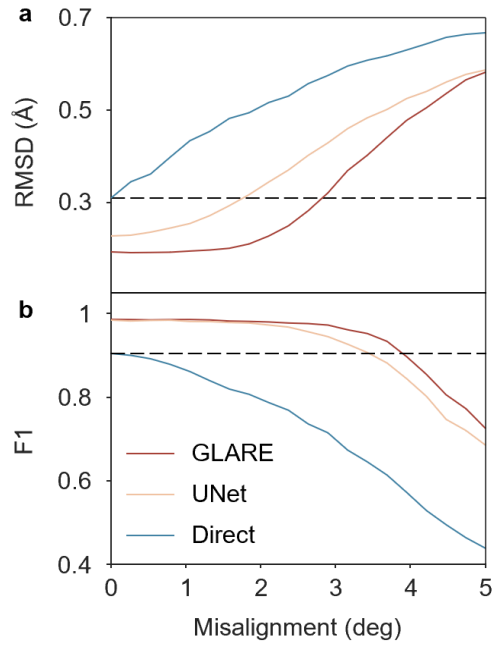

**Supplementary Figure 6 | GLARE performance on multi-slice simulated tilt series reconstruction. a**, RMSD of the atomic coordinates traced from the reconstructed volumes as a function of the misalignment introduced into the tilt-angle series. **b**, Corresponding F1 score as a function of misalignment. Atomic coordinates obtained from GLARE-refined volumes (red), UNet-refined volumes (gray), and directly traced raw reconstructed volumes (blue) are shown. The black dashed line indicates the baseline performance of direct atom tracing in the absence of misalignment.

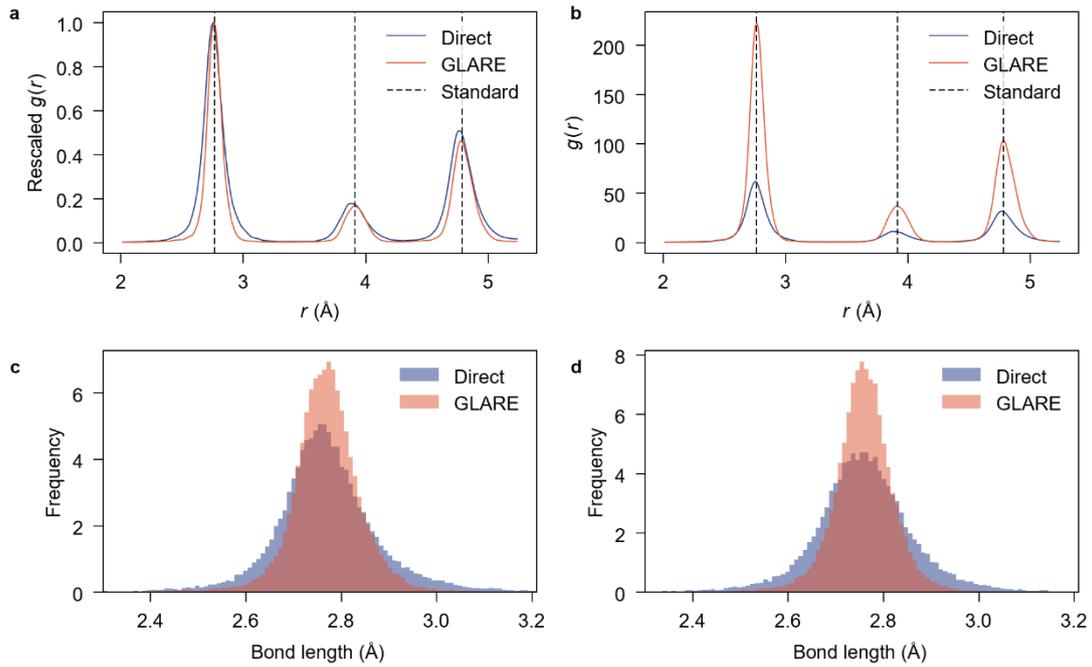

**Supplementary Figure 7 | PDF and bond length analyses of atomic structures obtained by GLARE and direct tracing. a,b**, PDF profiles rescaled by the first peak (**a**) and rescaled by the most distant point (**b**). Results from direct atom tracing are shown in blue, and results after GLARE refinement are shown in red. Dashed lines mark the reference PDF peak positions for a Pd@Pt nanoparticle, calculated using the mean values of Pd and Pt. **c,d**, Bond length distributions for the single-crystalline Pd@Pt nanoparticle (**c**) and the polycrystalline Pd@Pt nanoparticle (**d**). Blue histograms correspond to direct atom tracing, and red histograms correspond to GLARE-refined results.

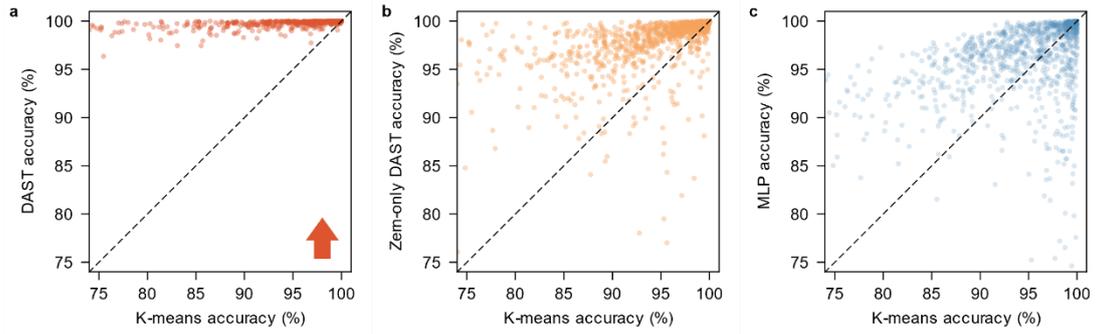

**Supplementary Figure 8 | Sample-by-sample accuracy comparison of different models against k-means clustering.** Scatter plots show the accuracies of (**a**) the DAST model, (**b**) the Zern-only DAST model, and (**c**) the MLP model, each plotted against the accuracy of k-means clustering. The black dashed line marks the $y = x$ reference; points above it indicate better performance than k-means.

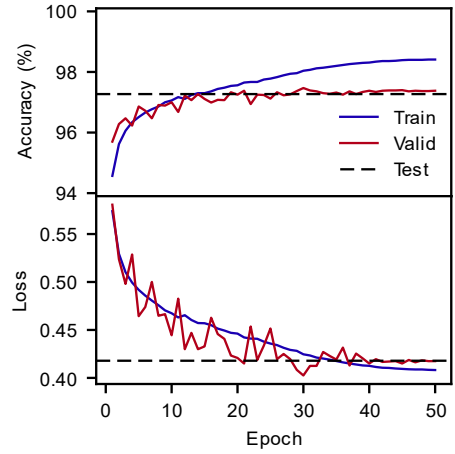

**Supplementary Figure 9 | Learning curves of the Zern-only DAST model.** Precision (top) and loss (bottom) are plotted as functions of training epoch. The blue curve represents the training set, the red curve the validation set, and the black dashed line indicates the test-set reference.

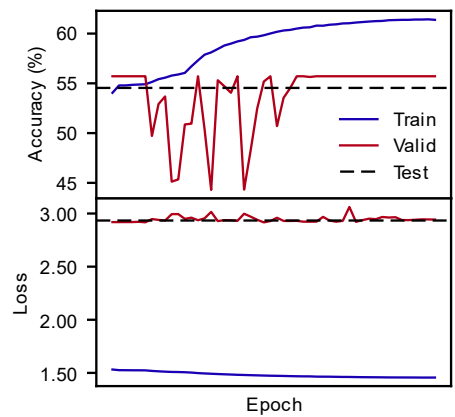

**Supplementary Figure 10 | Learning curves of the DAST model trained using only atomic coordinates.** Precision (top) and loss (bottom) are plotted against training epochs. The blue curve corresponds to the training set, the red curve to the validation set, and the black dashed line indicates the test-set reference.

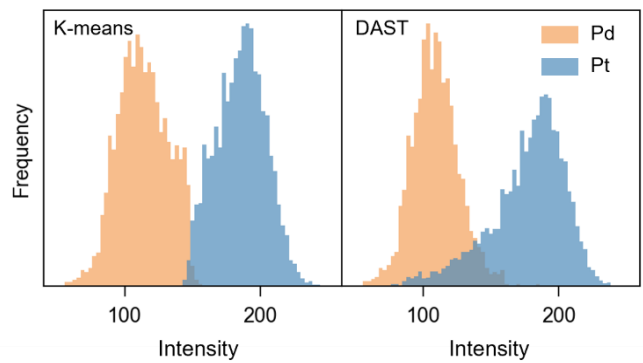

**Supplementary Fig. 11 | Atomic intensity distributions classified by k-means (left) and by the DAST model (right).**

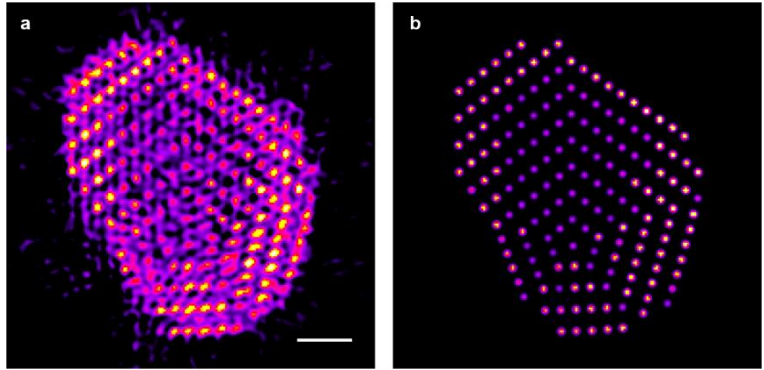

**Supplementary Figure 12 | GLARE refinement for low dose AET data.** 0.34-Å-thick slices of the raw reconstructed volume (**a**) and the GLARE refined volume (**b**) from 1/3-dose dataset. Scale bar, 1 nm.

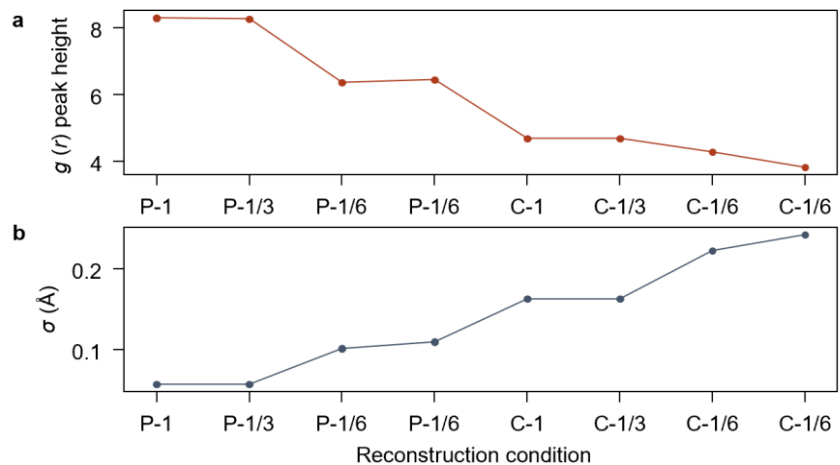

**Supplementary Figure 13 | PDF and bond length analysis of the atomic structures obtained using different workflows. a**, Height of the first PDF peak. **b**, Standard error of the bond length distribution.

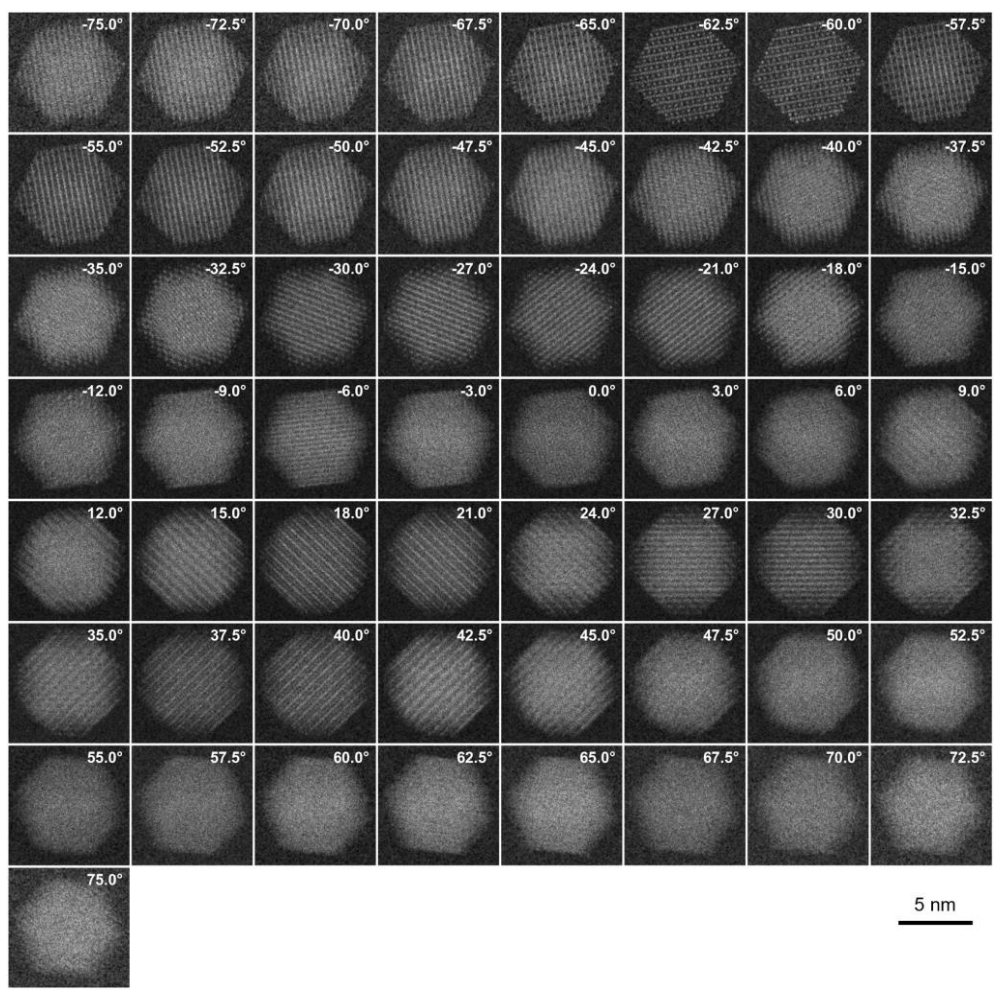

**Supplementary Figure 14 | Multi-slice–simulated low-dose tilt series of a CsPbBr$_3$ nanoparticle.** The tilt angles range from -75° to 75°, with an angular increment of 2.5° or 3°, giving a total of 57 projections. The tilt angle of each projection is indicated in the upper-right corner.

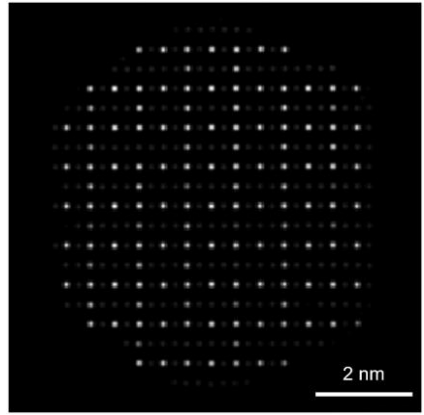

**Supplementary Figure 15 | A 0.34-Å-thick slice of the refined volume from the pre-trained GLARE model applied to the reconstructed CsPbBr$_3$ nanoparticle.**

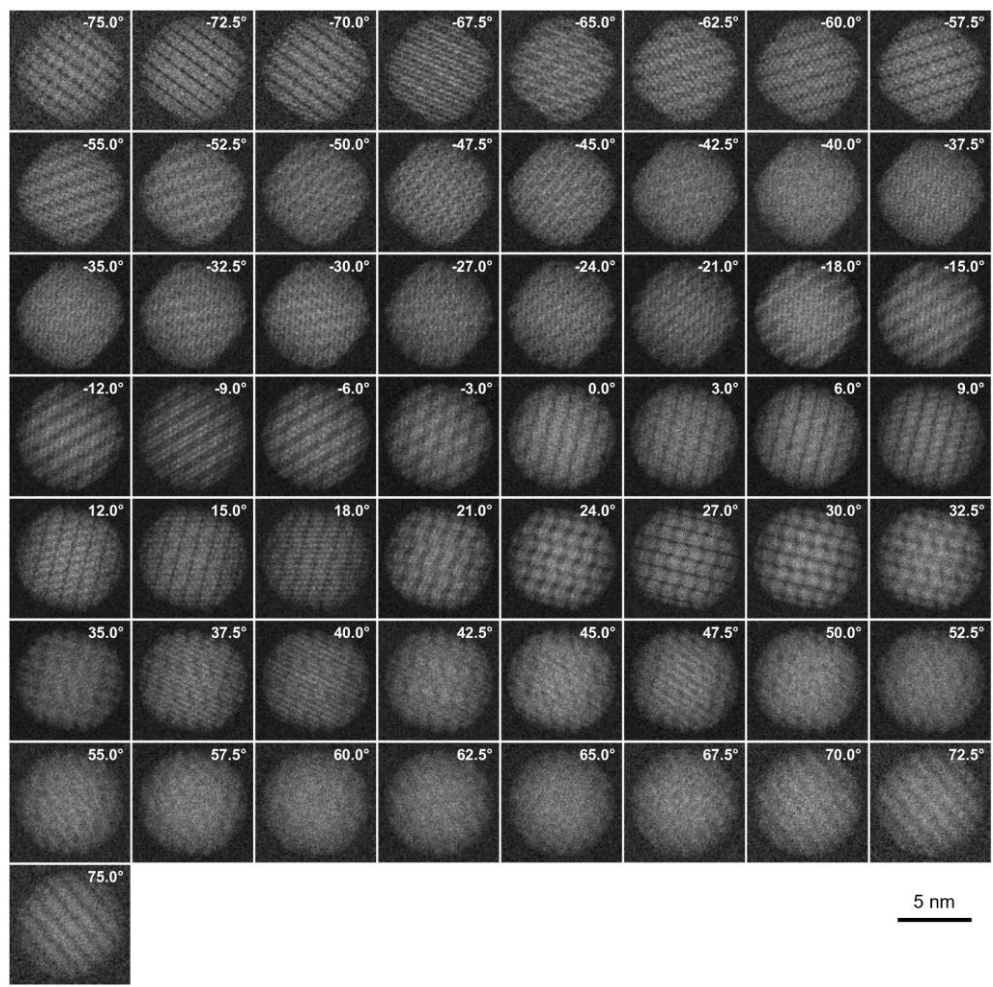

**Supplementary Figure 16 | Multi-slice–simulated low-dose tilt series of a ZSM-5 nanoparticle.**
The tilt angles range from -75° to 75°, with an angular increment of 2.5° or 3°, giving a total of 57 projections. The tilt angle of each projection is indicated in the upper-right corner.

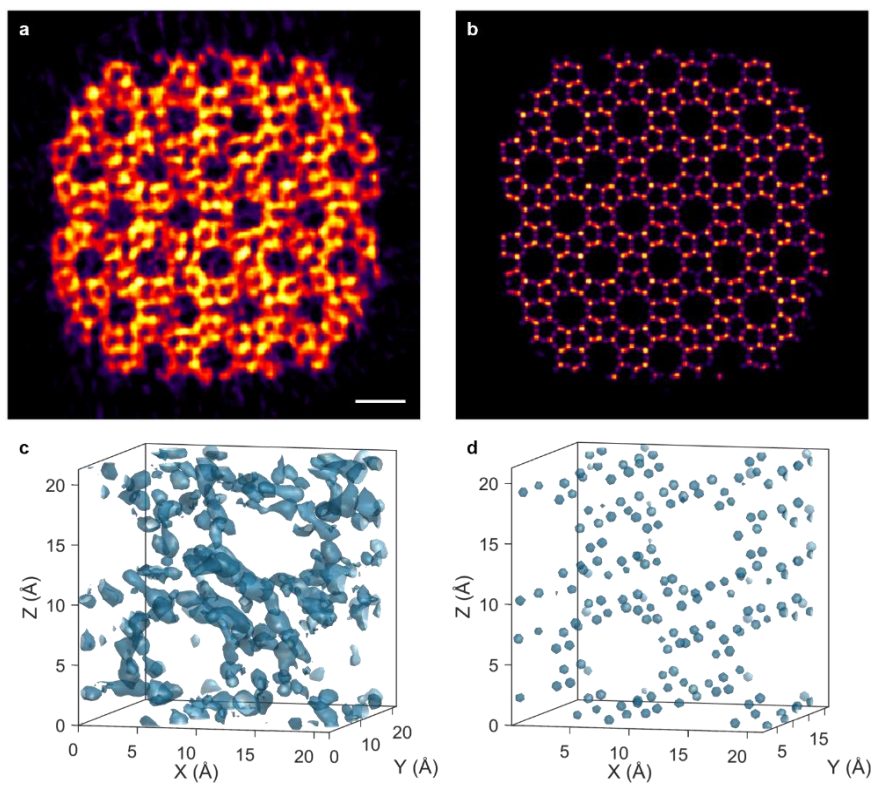

**Supplementary Figure 17 | Performance of the fine-tuned GLARE model on a ZSM-5 sample.**
**a**, A 4.1-Å-thick slice from the raw reconstructed volume. **b**, Corresponding slice from GLARE-refined volume. Scale bar, 1nm. **c**, Representative isosurface rendering of a region extracted from the raw reconstruction. **d**, Corresponding isosurface rendering from the refined volume.

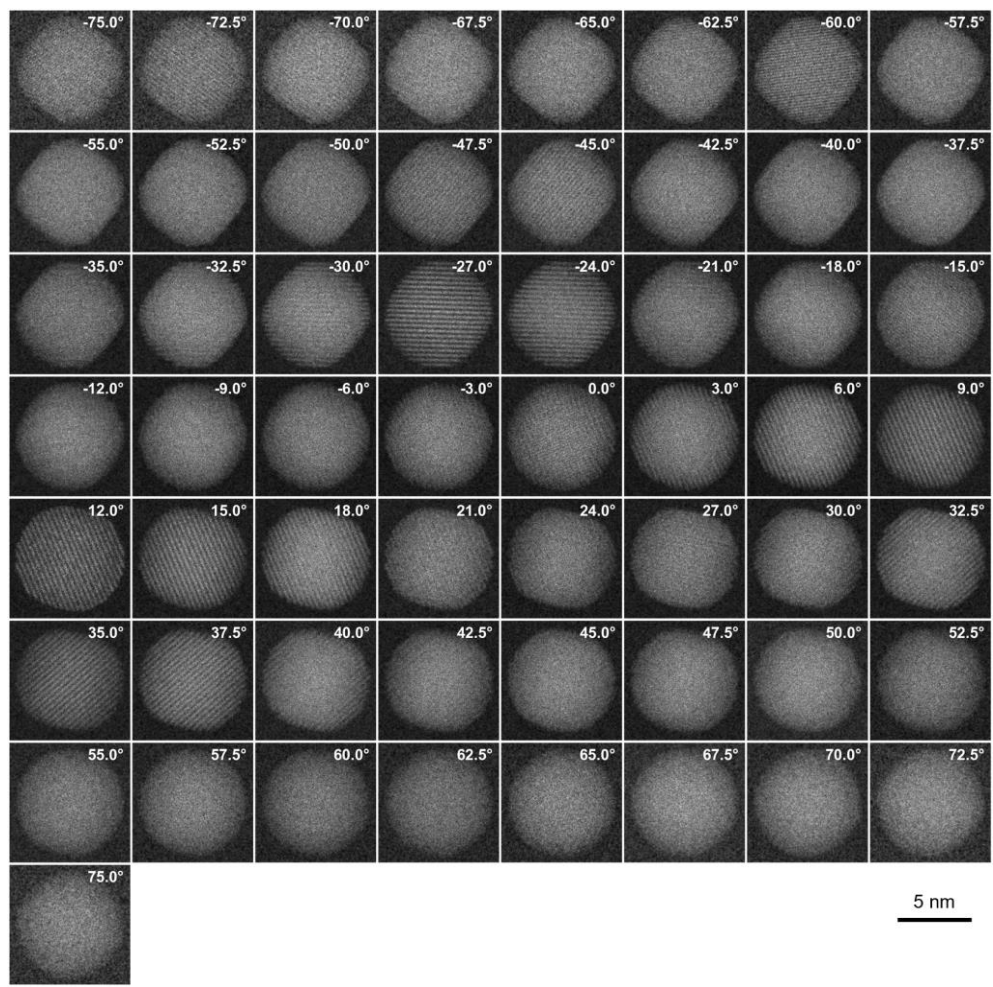

**Supplementary Figure 18 | Multi-slice–simulated low-dose tilt series of a CdSe nanoparticle.** The tilt angles range from -75° to 75°, with an angular increment of 2.5° or 3°, giving a total of 57 projections. The tilt angle of each projection is indicated in the upper-right corner.

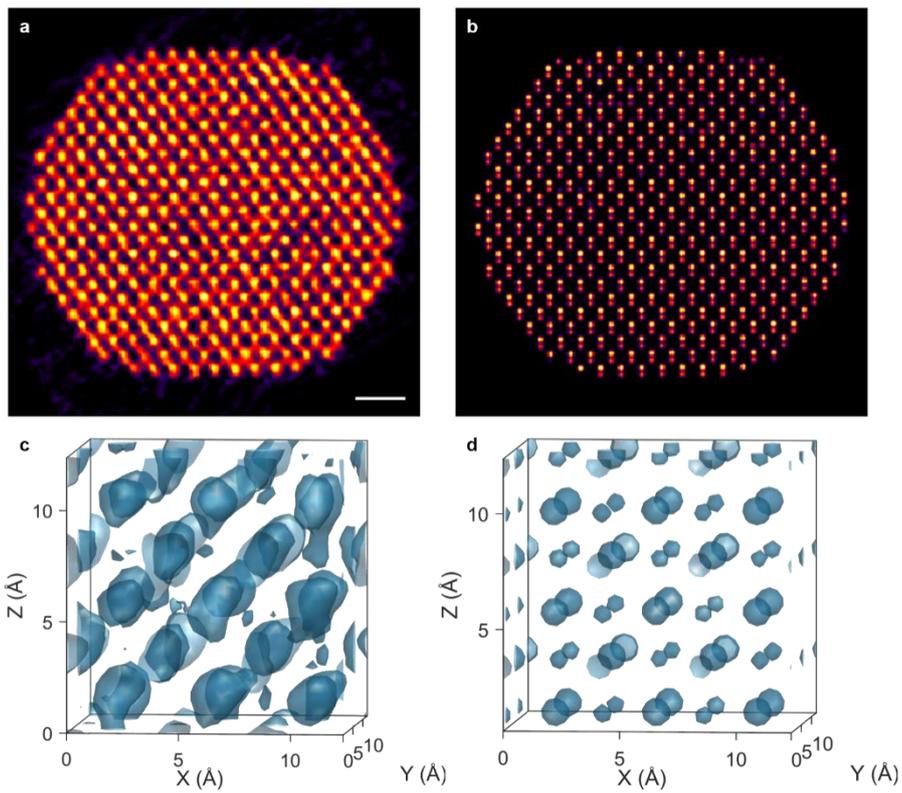

**Supplementary Figure 19 | Performance of the fine-tuned GLARE model on a CdSe sample.**
**a**, A 4.0-Å-thick slice from the raw reconstructed volume of the CdSe sample. **b**, Corresponding slice from the refined volume. Scale bar, 1nm. **c**, Representative isosurface rendering of a region extracted from the raw reconstruction. **d**, Corresponding isosurface rendering from the refined volume.

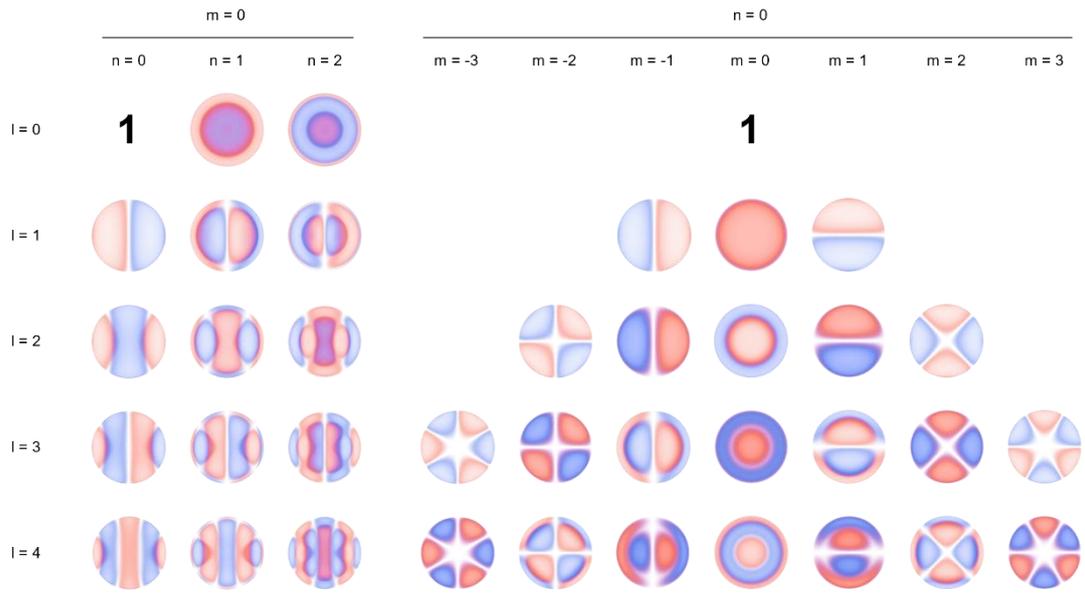

**Supplementary Figure 20 | Central slices of selected 3D Zernike polynomials.**